%% file: main.tex
\documentclass[sigplan, 10pt]{acmart}
\usepackage{subcaption}
\usepackage{listings}
\usepackage{ragged2e}
\usepackage{xcolor}
\usepackage{microtype}
\usepackage{tcolorbox}
\usepackage{macros}

\usepackage[linesnumbered,ruled,vlined]{algorithm2e}

\usepackage{mathabx}
\usepackage{booktabs}
\usepackage{balance}
\usepackage{adjustbox}
\usepackage{enumitem}
\usepackage{tabularx}
\usepackage{multirow} 
\usepackage{makecell}

\renewcommand\footnotetextcopyrightpermission[1]{}
\AtBeginDocument{%
  }

\setcopyright{none}              %

\makeatletter
\newcommand{\removelatexerror}{\let\@latex@error\@gobble}
\makeatother

\definecolor{mydarkred}{RGB}{200,0,0}

\SetCommentSty{mycommfont}
\definecolor{darkolivegreen}{rgb}{0.33, 0.42, 0.18}
\definecolor{cornellred}{rgb}{0.7, 0.11, 0.11}
\newcommand{\sys}{\textsc{Conduit}\xspace}

\makeatletter
\def\getTeXLiveYear#1 20#2#3#4#5\@nil{\def\TLYear{20#2#3}}
\expandafter\getTeXLiveYear\pdftexbanner\@nil
\ifnum\TLYear<2025
\else
    \patchcmd\caption@subtypehook{\let\label\subcaption@label}%
  {\let\label\subcaption@label\let\ltx@label\subcaption@label}{}{\fail}
\fi
\makeatother

\makeatletter
\AtBeginDocument{%
  \fancypagestyle{standardpagestyle}{%
    \fancyhf{}%
    \renewcommand{\headrulewidth}{\z@}%
    \renewcommand{\footrulewidth}{\z@}%
    \fancyfoot[C]{\if@ACM@printfolios\footnotesize\thepage\fi}%
  }%
  \pagestyle{standardpagestyle}%
}
\makeatother

\begin{document}

\title%
	[\sys: An Experience Data Plane for Distributed Reinforcement Learning]%
	{\sys: An Experience Data Plane\\for Distributed Reinforcement Learning}%

\author{Sitong Zhang}
\affiliation{%
	\institution{Aalto University}
}
\email{sitong.zhang@aalto.fi}

\author{Tuo Shi}
\affiliation{%
	\institution{Shenzhen University of Advanced Technology}
}
\email{shituo@suat-sz.edu.cn}

\author{Mario Di Francesco}
\affiliation{%
	\institution{Aalto University}
}
\email{mario.di.francesco@aalto.fi}

\author{Zeke Wang}
\affiliation{%
	\institution{Zhejiang University}
}
\email{wangzeke@zju.edu.cn}

\author{Bo Zhao}
\affiliation{%
	\institution{Aalto University}
}
\email{bo.zhao@aalto.fi}

\renewcommand{\shortauthors}{Zhang et al.}

\begin{abstract}
Distributed reinforcement learning (RL) scales training by parallelizing actors and learners around an \emph{Experience Buffer}. As RL workloads grow, however, the buffer becomes more than a replay queue: it is the storage substrate of a large-capacity, latency-critical \emph{experience path} that every iteration traverses to move, transform, sample, and batch experiences before learner updates can begin. Existing RL systems embed this path inside framework control flow or expose it as a request-driven buffer service, leaving experience placement fixed and experience-path work difficult to schedule independently as a runtime-level optimization target. 
	
	We present \textbf{\sys}, a framework-agnostic runtime that exposes RL experience management as an explicit systems optimization problem. At its core is the \emph{Experience Data Plane} (EDP), a runtime abstraction that separates RL experience-handling semantics from framework-specific execution logic by exposing \emph{experience ingestion}, \emph{experience placement}, and \emph{experience delivery} as explicit control points. Built on EDP, \sys introduces \emph{capacity-constrained, bandwidth-aware placement}, which distributes experience state across CPU/GPU memory tiers and nodes under heterogeneous interconnect and device-memory constraints, and \emph{latency-aware scheduling}, which controls when experience-path handling runs to reduce exposed experience-path latency while preserving RL semantics. Integrated with RLlib without changing its framework execution logic, \sys reduces exposed experience-path latency by up to 97\% and end-to-end iteration latency by up to 38\%, scales to 1{,}024 GPUs, and preserves convergence.\looseness=-1
\end{abstract}

\settopmatter{authorsperrow=3}
\settopmatter{printfolios=true}
\maketitle
\hypersetup{pdftitle={Conduit: An Experience Data Plane for Distributed Reinforcement Learning}}

\input{sections/introduction}

\input{sections/background}
\input{sections/design}

\input{sections/decoupling}
\input{sections/placement}

\input{sections/scheduling}

\input{sections/evaluation}
\input{sections/related}

\balance
\bibliographystyle{ACM-Reference-Format}
\bibliography{main}

\pagebreak
\nobalance
\appendix
\input{sections/appendix}

\end{document}

%% file: sections/introduction.tex
\section{Introduction}
\label{sec:intro}

Distributed reinforcement learning (RL) increasingly operates at scale, with large observations and heavy per-sample payloads
~\cite{huang2022distributed,zhang2022pipo,chunduri2022zeus,jasny2020db4ml,yan2023join,DBLP:journals/corr/abs-2510-05943,DBLP:conf/iclr/ChenFZG00HCS25,Jern2025AgentQFL,yao2023deepspeed,sheng2025hybridflow,fu2025areal}.
In a typical distributed RL system, actors generate experiences, learners update models, and an \emph{Experience Buffer} bridges the two~\cite{zhu2023msrl,liang2021rllib,DBLP:conf/osdi/MoritzNWTLLEYPJ18,mei2023srl}.
As workloads grow%
, however, this buffer %
becomes the storage substrate of a large-capacity, latency-critical \emph{experience path}.
Modern experiences carry large visual, multimodal, or long-context payloads, and each iteration repeatedly moves, transforms, samples, and batches these heavy payloads before learner update can begin.

\begin{figure}[t]
	\centering
	\includegraphics[width=\linewidth]{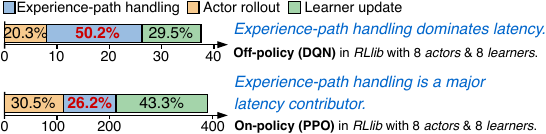}
	\caption{Experience-path latency is a major component of iteration time in distributed RL: 50.2\% in off-policy Deep Q-Network (DQN) and 26.2\% in on-policy proximal policy optimization (PPO) across representative workloads on an 8$\times$A100 cluster.}
	\label{fig:breakdown}
\end{figure}

\begin{figure}[t]
	\centering
	\includegraphics[width=\linewidth]{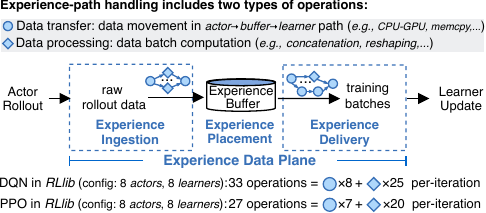}
	\caption{Experience-path handling drives latency through transfer and processing operations. Existing systems embed this handling in their actor/learner workflow; \sys instead abstracts it into an explicit \emph{Experience Data Plane}. 
	}
	\label{fig:edp_overview}
\end{figure}
 
Despite extensive effort to optimize actors and learners, distributed RL systems still overlook \emph{experience-path latency}: the repeated work that converts raw rollouts into training-ready batches, i.e., \emph{experience-path handling}.
\F\ref{fig:breakdown} shows that this latency accounts for 50.2\% of iteration time in off-policy DQN~\cite{mnih2013playing} and 26.2\% in on-policy PPO~\cite{schulman2017proximal}. 
\F\ref{fig:edp_overview} shows where this latency comes from: experience-path handling runs dozens of transfer and processing operations every iteration (33 for DQN, 27 for PPO in {RLlib}).

The experience path is %
also \emph{capacity-constrained}: large buffers often exceed a single GPU and must be distributed across CPUs, GPUs, and nodes on heterogeneous fabrics. %
Existing systems, however, treat capacity management, placement, and scheduling as local %
framework tactics rather than as %
a runtime problem.

Experience buffers built into RL frameworks such as RLlib~\cite{liang2021rllib}, MSRL~\cite{zhu2023msrl}, and SRL~\cite{mei2023srl} embed handling within actor/learner execution at fixed locations (see \F\ref{fig:edp_overview} for the RLlib case); service-based systems %
such as Reverb~\cite{cassirer2021reverb} and Gear~\cite{wang2023gear} are driven by incoming requests rather than proactively scheduling critical-path work to reduce end-to-end latency. %
Both designs treat the experience buffer as an internal framework component or service endpoint, rather than exposing it as a runtime interface for optimization.

Our key insight is that distributed RL needs precisely this missing abstraction: an explicit runtime interface for managing and optimizing experience data, which we call the \emph{Experience Data Plane} (EDP). %
RL frameworks retain control over \emph{which} experiences to store and consume, while EDP manages the systems-level decisions of \emph{where} those experiences reside and \emph{when} ingestion and delivery occur.

We introduce \sys to realize EDP as a framework-agnostic experience-buffer runtime that exposes experience ingestion, placement, and delivery as explicit system-level control points (\F\ref{fig:edp_overview}) while preserving framework execution logic. %
Built on EDP, \sys introduces two mechanisms: \emph{capacity-constrained, bandwidth-aware placement}, which distributes the Experience Buffer across CPU and GPU memory under heterogeneous interconnect bandwidths and device-memory limits; and \emph{latency-aware scheduling}, which determines when ingestion and delivery run. \looseness=-1 

\sys makes the following contributions:

\tinyskip
\mypar{(1) Experience Data Plane as a runtime abstraction}
We introduce the \emph{Experience Data Plane} (EDP), a runtime abstraction that exposes \emph{experience ingestion}, \emph{experience placement}, and \emph{experience delivery} as explicit control points rather than framework-internal side effects.
EDP decouples RL experience semantics from framework-specific execution logic, creating an optimization surface for data residency, capacity management, and scheduling in distributed RL. (\S\ref{sec:decouple})

\tinyskip
\mypar{(2) Capacity-constrained, bandwidth-aware placement}
We design a placement mechanism that distributes the Experience Buffer across CPU/GPU memory tiers and nodes by jointly accounting for actor\,$\rightarrow$\,buffer\,$\rightarrow$\,learner path costs, heterogeneous interconnect bandwidths, and device-memory constraints.
This allows \sys to support buffers larger than a single GPU while reducing data-movement cost on heterogeneous fabrics. (\S\ref{sec:placement})

\tinyskip
\mypar{(3) Latency-aware scheduling of experience-path handling}
We design a scheduling mechanism that decides when experience ingestion and delivery run, and at what granularity, to reduce exposed experience-path latency while preserving on-policy freshness and off-policy replay semantics.
This allows \sys to overlap experience-path handling with rollout and learner update when semantics permit, while amortizing overhead through batching and pipelining. (\S\ref{sec:scheduling})

\smallskip
We implement \sys and integrate it with RLlib~\cite{liang2021rllib}---a state-of-the-art reinforcement-learning framework---without changing its execution logic.
We further show that \sys generalizes beyond RLlib: it integrates into {SRL}~\cite{mei2023srl}, a distributed RL framework with a different, streaming-dataflow execution model (Appendix~\ref{app:srl}), and extends to large language model (LLM) post-training with the \texttt{Verl}~\cite{sheng2025hybridflow} framework (Appendix~\ref{app:llm-post-training}).
Across on-policy and off-policy RL workloads, \sys reduces both exposed experience-path latency (up to 97\%) and end-to-end iteration latency (up to 38\%), and scales efficiently to 1{,}024 GPUs.
These results establish EDP as a practical optimization surface for distributed RL: a runtime abstraction that decouples capacity, placement, and scheduling decisions from framework execution logic.

%% file: sections/background.tex
\section{Background and Motivation}
\label{sec:motivation}

This section motivates treating experience-path handling as an independently optimizable runtime component on the actor\,$\rightarrow$\,Experience Buffer\,$\rightarrow$\,learner path.
\S\ref{subsec:ddrl} frames distributed RL through this path, \S\ref{subsec:roles} summarizes its responsibilities, \S\ref{subsec:buffer_req} states the three challenges that motivate \sys, and \S\ref{subsec:limitations} reviews existing approaches.

\subsection{Actor\,$\rightarrow$\,Experience Buffer\,$\rightarrow$\,Learner Data Path}
\label{subsec:ddrl}

In RL, an agent learns a policy (typically a deep neural network, DNN) to act in an environment~\cite{sutton2018reinforcement}.
Each training iteration executes the current policy to generate \emph{rollouts} (experience tuples such as state, action, reward, and next state) and then performs \emph{learner update} to train the policy DNN on those experiences.

\begin{figure}[t]
	\centering
	\includegraphics[width=0.99\linewidth]{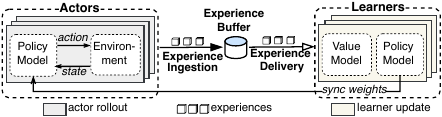}
	\caption{Distributed deep reinforcement learning architecture.}
	\label{fig:ddrl}
\end{figure}

In practice, these tuples are rarely flat records: modern experiences increasingly contain large visual, multimodal, or long-context payloads augmented with termination flags and auxiliary metadata (\eg \code{policy\_info}, \code{episode\_ID})~\cite{tfagents_trajectory,rllib_episodes}.
To scale training, modern RL systems parallelize actor rollout and learner update~\cite{liu2024acceleration} by running many actors and learners, bridged by an \emph{Experience Buffer} that ingests rollouts and serves training-ready batches (\F\ref{fig:ddrl}).

The \emph{actor\,$\rightarrow$\,Experience Buffer\,$\rightarrow$\,learner} connection forms a large-capacity, latency-critical experience path.
Every iteration must repeatedly move, transform, sample, and batch these heavy payloads into training-ready data under strict algorithmic constraints (\eg freshness requirements in on-policy RL and reuse via replay in off-policy RL). 
To understand how capacity demands and execution latency along this path shape end-to-end training efficiency, the next subsection characterizes its logical responsibilities and connects them to the underlying operator graph. \looseness=-1

\subsection{Experience Path Responsibilities}
\label{subsec:roles}

Experience-path handling spans two complementary levels:
(i) a logical view that defines \emph{what} the path provides (\emph{ingestion}, \emph{delivery}, and \emph{placement}), and
(ii) a physical execution plan that defines \emph{how} this logical view is realized as operator graphs (transfer and processing operators).

\begin{table*}[t]
	\renewcommand{\arraystretch}{1}
	\small
	\centering
	\caption{Comparison of existing experience buffer implementations for distributed RL.}
	\label{tab:existing_buffers}
	\resizebox{\linewidth}{!}{%
	\begin{tabular}{lllll}
		\toprule
		\textbf{Category} & \textbf{System} & \textbf{Decoupling} & \textbf{Placement} & \textbf{Scheduling} \\ \hline
		
		\multirow{4}{*}{\makecell[l]{\textbf{Built-in}\\\textbf{Experience Buffers}}}
		& RLlib~\cite{liang2021rllib} & \multirow{4}{*}{\makecell[l]{Embedded module}} & Fixed (CPU) &
		\multirow{2}{*}{\makecell[l]{Reactive (inline in actor/learner loops)}} \\ \cline{2-2} \cline{4-4}
		& MSRL~\cite{zhu2023msrl} & & Fixed (CPU/GPU) & \\ \cline{2-2} \cline{4-5}
		& \multirow{2}{*}{SRL~\cite{mei2023srl}} & & \multirow{2}{*}{Fixed (pinned CPU)} &
		\multirow{2}{*}{\makecell[l]{Reactive (fixed one-step overlap)}} \\
		& & & & \\ \hline
		
		\multirow{2}{*}{\makecell[l]{\textbf{Experience Buffer}\\\textbf{Services}}}
		& Reverb~\cite{cassirer2021reverb} & \multirow{2}{*}{\makecell[l]{Process-decoupled service}} & Fixed (CPU) &
		\multirow{2}{*}{\makecell[l]{Reactive (request-driven)}} \\ \cline{2-2} \cline{4-4}
		& Gear~\cite{wang2023gear} & & Fixed (pinned CPU \& GPU‑driven access) & \\ \hline
		
		\midrule

\multirow{2}{*}{\makecell[l]{\textbf{Experience}\\\textbf{Management Layer}}}
& \multirow{2}{*}{\textbf{\sys}} 
& \multirow{2}{*}{\makecell[l]{\text{Multi-dimensional Decoupling} \\ (system, algorithm, hardware)}} 
& \multirow{2}{*}{\makecell[l]{\text{Adaptive} \\ (capacity-constrained, bandwidth-aware)}} 
& \multirow{2}{*}{\makecell[l]{\text{Proactive} (schedulable execution  \\ timing \& processing granularity)}} \\ 
& & & & \\

		\bottomrule
	\end{tabular}%
	}%
\end{table*}

\myparr{(i) Logical view}
comprises two functional operations, \emph{experience ingestion} and \emph{experience delivery}, and one physical control point, \emph{experience placement}.

\textbf{Experience ingestion} collects actor rollouts into buffer-resident experiences by running data-plane operators that materialize and organize incoming records.
In on-policy training, ingestion assembles step-level tuples into trajectories and applies preprocessing~\cite{neves2024advances,schulman2017proximal} (\eg boundary detection, padding/truncation, normalization, and computing returns/advantages) before forming batches for the current iteration.
In off-policy training, it inserts transitions into a persistent replay dataset and maintains indices and metadata %
for future sampling and eviction~\cite{horgan2018distributed,schaul2015prioritized}.

\textbf{Experience delivery} extracts experiences from the buffer and produces training-ready batches for the learner by running operators that \emph{select}, \emph{assemble}, and \emph{transform} records under algorithmic constraints~\cite{schaul2015prioritized,horgan2018distributed}.
In on-policy training, delivery consumes freshly ingested experiences within the same iteration by partitioning trajectories into mini-batches and applying lightweight batch transformations (\eg shuffling and padding~\cite{schulman2017proximal}).
In off-policy training, it samples from a persistent replay dataset, materializes the selected transitions, and performs batch/layout transformations (\eg concatenation and reshaping) required by the learner~\cite{mnih2015human}.

\textbf{Experience placement} controls \emph{where} and \emph{how} experience state resides across distributed resources (\eg CPU memory, pinned host memory, GPU memory, or distributed across multiple nodes).
As a logical knob, placement does not change the RL semantics of \emph{what} experience ingestion or delivery must accomplish, but it dictates how capacity is distributed across devices.
It determines where data-plane operators \emph{read/write} records and which transfer edges are exercised. Because every experience must traverse this path, placement directly governs whether large datasets can fit in available memory, as well as the cost of cross-device data movement on heterogeneous fabrics.

\myparr{(ii) Physical execution}
realizes experience ingestion and delivery as operator graphs with two operator types:
\emph{data transfer} moves rollouts/batches across devices, memory domains, and nodes, while \emph{data processing} transforms them into learner-consumable layouts.
Transfer operators include CPU$\leftrightarrow$GPU copies, GPU$\leftrightarrow$GPU transfers, inter-node send/recv, and shard movement via collectives~\cite{de2024exploring,vadhiyar2000automatically}.
Processing operators include trajectory assembly, concat/ stack, reshape/ slice, sampling, shuffling, and batching, optionally with padding/ packing for variable-length sequences~\cite{sutton2018reinforcement}.

Thus, ingestion and delivery define the logical phases, transfer and processing operators define physical execution, and experience placement determines capacity and locality, shaping the dominant transfer costs on the actor\,$\rightarrow$\,Experience Buffer\,$\rightarrow$\,learner path in \F\ref{fig:ddrl}.
 
\subsection{Experience-Path Challenges and Opportunities}
\label{subsec:buffer_req}
Existing RL frameworks face three challenges in managing the experience path.
Each challenge motivates the design of the \emph{Experience Data Plane} (EDP), which turns it into an optimization opportunity.

\begin{figure}[t!]
	\centering
	\includegraphics[width=1.02\linewidth]{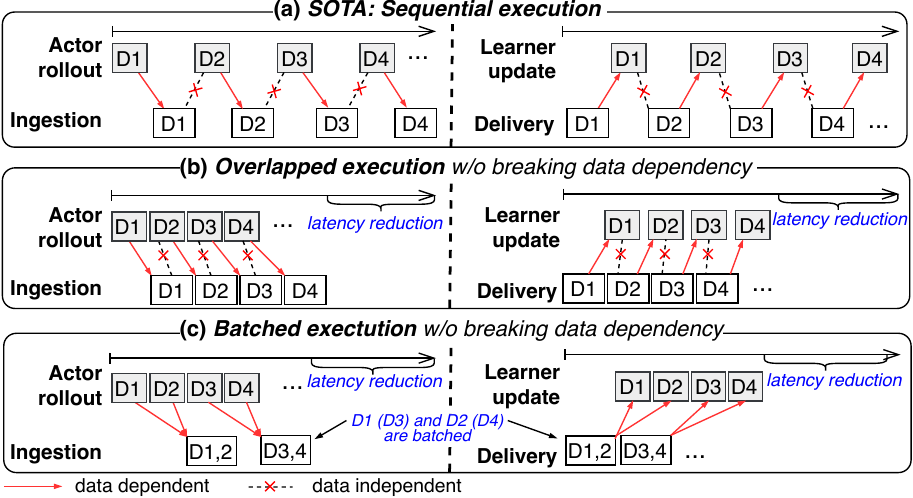}
	\caption{Operator dependencies (\code{D1}--\code{D4} are batches): (a) sequential, (b) overlapped via \emph{timing}, (c) overlapped and batched via joint \emph{timing} and \emph{granularity}.}
	\label{fig:dependency}
\end{figure}

\mypar{Challenge 1: Coupled execution serializes handling onto the critical path}
Most distributed RL frameworks embed ingestion and delivery inside actor/learner execution: ingestion is tied to actor rollout, and delivery is tied to learner update.
Although dependencies only constrain operators \emph{within} a batch, hard-wiring these operator graphs into fixed execution loops turns them into global barriers that block pipelining across batches (\eg ingesting \code{D$_i$} while generating \code{D$_{i+1}$}, or delivering \code{D$_{i+1}$} while updating on \code{D$_i$}; \F\ref{fig:dependency}), forcing handling latency onto the critical path.

\mypar{Opportunity: Decoupling} By \emph{decoupling} ingestion and delivery from framework execution, the EDP turns them into first-class objects that the system controls directly, granting three forms of autonomy:
\emph{execution autonomy}, to run them asynchronously and pipeline across batches when dependencies allow;
\emph{semantic autonomy}, to serve on-policy freshness and off-policy replay under one interface; and
\emph{hardware autonomy}, to detach experience placement from fixed framework bindings (\eg CPU or a specific GPU).
 
\begin{figure}[t!]
	\centering
	\begin{subfigure}[t]{0.45\linewidth}
		\centering
		\includegraphics[width=\linewidth]{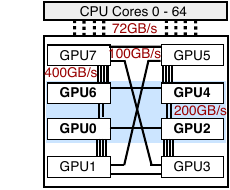}
		\caption{Non-uniform GPU fabric.}
		\label{fig:fabric_topo}
	\end{subfigure}
	\hfill
	\begin{subfigure}[t]{0.50\linewidth}
		\centering
		\includegraphics[width=\linewidth]{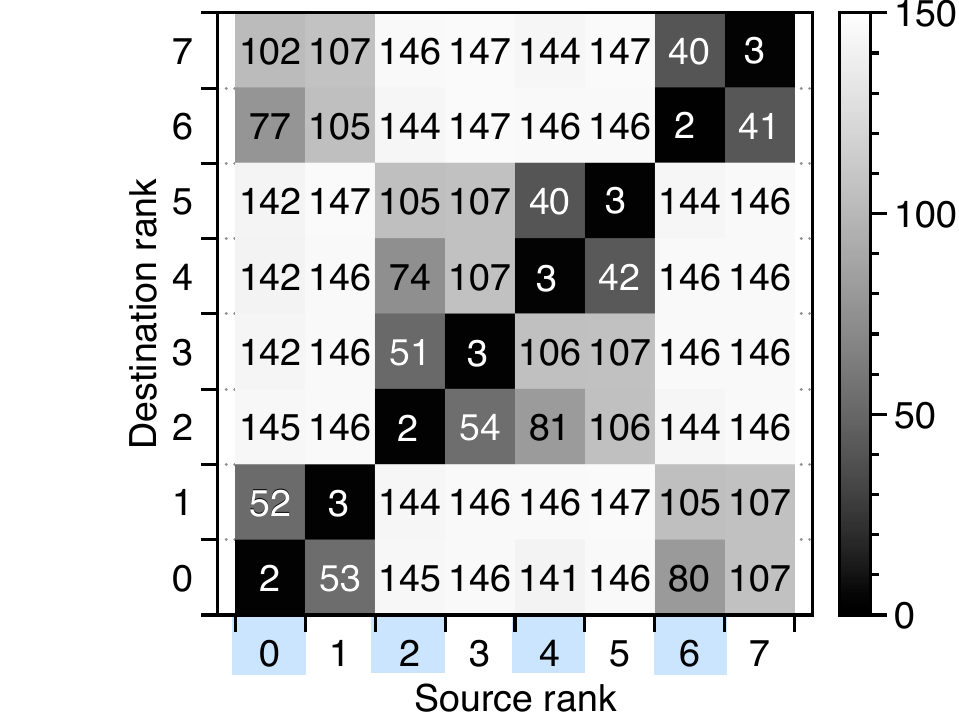}
		\caption{GPU$\leftrightarrow$GPU latency heatmap.}
		\label{fig:fabric_heatmap}
	\end{subfigure}

	\caption{Non-uniform intra-node transfers on an 8-GPU AMD MI250X node. (a) interconnect topology; (b) GPU$\leftrightarrow$GPU latency variation for a 5\,GB transfer, making placement performance-critical.}
	\label{fig:nonuniform_fabric}
\end{figure}

\tinyskip
\mypar{Challenge 2: Fixed placement either overflows memory or pays for slow fabric transfers}
Where experience data resides determines both capacity and transfer cost, and the two pull against each other.
Capacity binds at scale: in visual RL with 128$\times$128--256$\times$256 RGB inputs~\cite{gu2023maniskill2,o2024open,hansen2024tdmpc2}, a $10^6$-sample buffer reaches 48--192\,GB, exceeding a single GPU.
But spilling elsewhere can be costly. \F\ref{fig:nonuniform_fabric} illustrates this non-uniformity within a single node\footnote{We profile these fabrics on an AMD MI250X supercomputer}: GPU pairs differ in bandwidth (100--400\,GB/s) and CPU$\leftrightarrow$GPU is slower (72\,GB/s) (\F\ref{fig:fabric_topo}), so a 5\,GB GPU$\leftrightarrow$GPU transfer spans 40--147\,ms, a 3.7$\times$ spread (\F\ref{fig:fabric_heatmap}).
A fixed choice (always CPU or a fixed GPU) thus either overflows memory or repeatedly pays for slow transfers.

\mypar{Opportunity: Placement} The EDP's hardware autonomy lets \sys treat placement as a runtime decision rather than a fixed binding.
Through \emph{capacity-constrained, bandwidth-aware placement}, \sys distributes the buffer across CPU, GPU, and node tiers to fit memory limits while steering data onto fast edges---scaling capacity beyond a single device and avoiding the slowest fabric edges at the same time.

\tinyskip
\mypar{Challenge 3: Synchronous handling fully exposes its latency and wastes per-invocation overhead}
Experience handling is far from free: ingestion and delivery perform non-trivial processing (trajectory assembly, batching, sampling, reshaping, shuffling, padding).
Run synchronously between rollout and update, this work lands entirely on the critical path, where it becomes a major component of iteration time (\F\ref{fig:breakdown}); executed one unit at a time, its fixed per-invocation overhead is also paid repeatedly.

\begin{figure}[t!]
	\centering
	\includegraphics[width=0.7\linewidth]{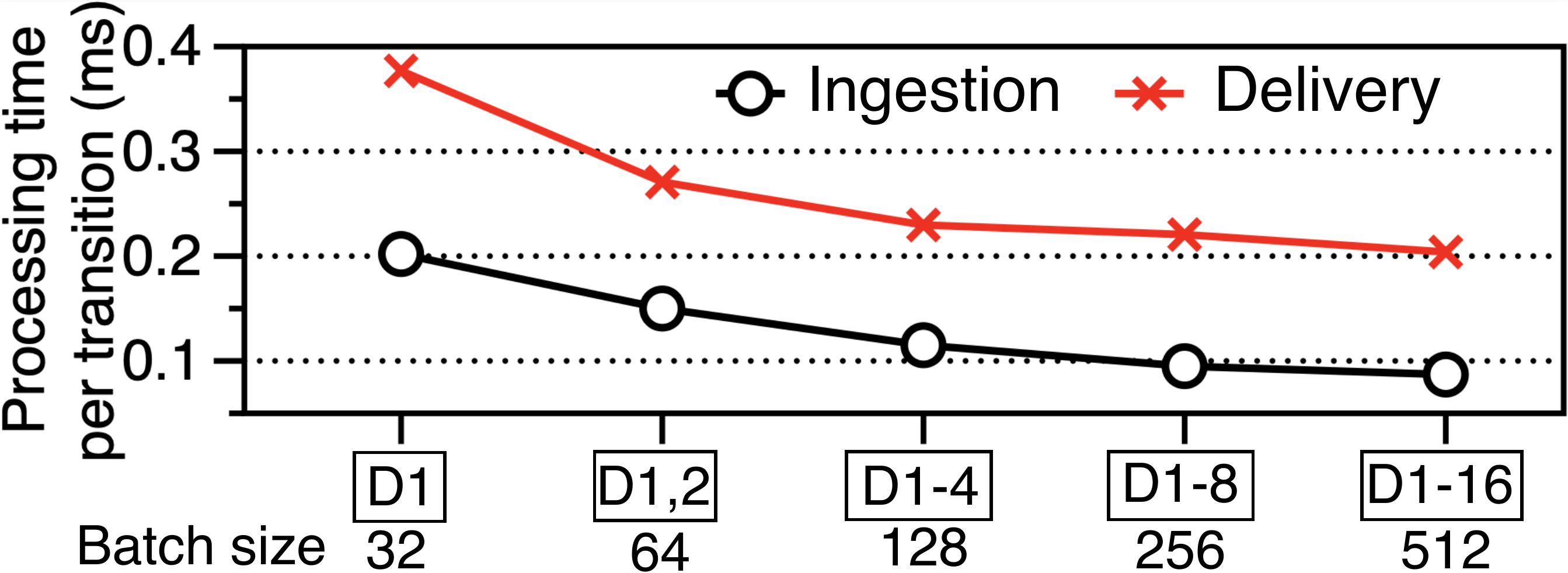}
	\caption{Larger operator batches reduce per-transition processing cost for both ingestion and delivery.}
	\label{fig:batch_process}
\end{figure}

\mypar{Opportunity: Scheduling} The EDP's execution autonomy lets \sys schedule experience-path handling instead of running it inline.
Because dependencies bind operators only \emph{within} a batch (\F\ref{fig:dependency}), two knobs reduce cost.
\emph{Timing} overlaps ingestion for \code{D$_i$} with rollout for \code{D$_{i+1}$}, and delivery for \code{D$_{i+1}$} with the update on \code{D$_i$} (\F\ref{fig:dependency}b); \emph{granularity} batches trajectories per invocation to amortize overhead (\F\ref{fig:dependency}c).
Both help: on RLlib~\cite{liang2021rllib}, increasing the operator batch from 32 to 512 transitions cuts amortized ingestion from 0.20 to 0.085\,ms/transition (2.4$\times$) and delivery from 0.38 to 0.21\,ms (1.8$\times$) (\F\ref{fig:batch_process}).

\tinyskip
\mypar{Summary}
Together, addressing these three challenges yields \sys's optimization opportunities---\emph{decoupling}, \emph{placement}, and \emph{scheduling}---which it targets end-to-end on the actor\,$\rightarrow$\,Experience Buffer\,$\rightarrow$\,learner data path.

\subsection{Limitations of Existing RL Experience Buffers}
\label{subsec:limitations}

Existing RL buffer designs fall into two structural categories: framework-internal modules and standalone service endpoints (\T\ref{tab:existing_buffers}). In both cases, experience handling remains reactive, placement-fixed, and tightly bound to framework execution, preventing systematic optimization across decoupling, placement, and scheduling.

\mypar{Built-in experience buffers (embedded and reactive)}
RL frameworks such as RLlib~\cite{liang2021rllib} and SRL~\cite{mei2023srl} implement the experience buffer as an internal framework component, so experience ingestion and delivery execute inline with actor rollout and learner update.
This hard-wires experience handling into framework-specific execution logic, preventing principled pipelining across batches (see \F\ref{fig:dependency}).
Built-in experience buffers are also placement-fixed (CPU or pinned host memory), failing to manage capacity or optimize transfer costs across distributed GPU resources.

MSRL~\cite{zhu2023msrl} improves task placement across actors and learners via a fragmented dataflow graph, which is complementary. However, it does not expose an explicit runtime surface for the experience path itself, leaving experience handling reactive and tied to task execution.

\tinyskip
\mypar{Experience buffer services (throughput-scaled, but not latency-optimized)}
Reverb~\cite{cassirer2021reverb} and Gear~\cite{wang2023gear} decouple the experience buffer at the process level.
However, they treat the buffer as a service endpoint rather than as a runtime control surface. They retain fixed placement (CPU resident) and request-driven execution.
Reverb stores experience data in pageable CPU memory. Gear stores it in pinned CPU memory while using GPUs to leverage DMA and RDMA %
reads, rather than dynamically placing experience data across GPU boundaries.
By ignoring the joint optimization of capacity-constrained placement and proactive scheduling, these systems focus on insertion/sampling throughput but do not directly optimize exposed experience-path latency.

\tinyskip
\mypar{Summary}
Prior work treats the experience buffer as an internal component or external service rather than as an explicit Experience Data Plane, motivating \sys.

%% file: sections/design.tex
\section{Overview of \sys}
\label{sec:overview}

\sys is a runtime system that realizes the \emph{Experience Data Plane} (EDP) for the large-capacity, latency-critical experience path.
It separates RL experience handling from framework-specific execution by exposing \emph{experience ingestion}, \emph{experience placement}, and \emph{experience delivery} as runtime control points.
\sys integrates with existing distributed RL frameworks by intercepting buffer interfaces; on-policy freshness and off-policy replay semantics bound feasible scheduling and placement, so \sys optimizes capacity and latency within these bounds.

\begin{figure}[t]
	\centering
	\includegraphics[width=0.98\linewidth]{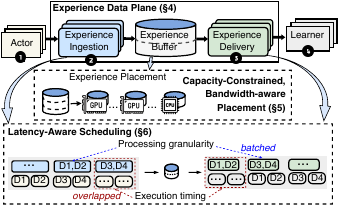}
	\caption{\sys workflow: the \emph{Experience Data Plane} exposes three control points---\emph{experience ingestion}, \emph{placement}, and \emph{delivery}---around the Experience Buffer.}
	\label{fig:bufferservice}
\end{figure}

As shown in \F\ref{fig:bufferservice}, the EDP exposes three control points around the persistent \emph{Experience Buffer}---\emph{experience ingestion}, \emph{placement}, and \emph{delivery}---refining the rigid actor\,$\rightarrow$\,Experience Buffer\,$\rightarrow$\,learner execution into a fine-grained, independently controllable experience path.
Actors push raw rollout payloads asynchronously (\myc{1}); \sys ingests them into the Experience Buffer (\myc{2}) and delivers training-ready batches (\myc{3}); learners then pull these staged batches for model updates (\myc{4}).
Because ingestion and delivery run as operators decoupled from actor and learner execution, they become \emph{pipelineable constraints}: \sys can overlap ingestion with actor-environment interaction and delivery with policy DNN training, reducing exposed experience-path latency.

\begin{figure}[t]
	\centering
	\includegraphics[width=0.98\linewidth]{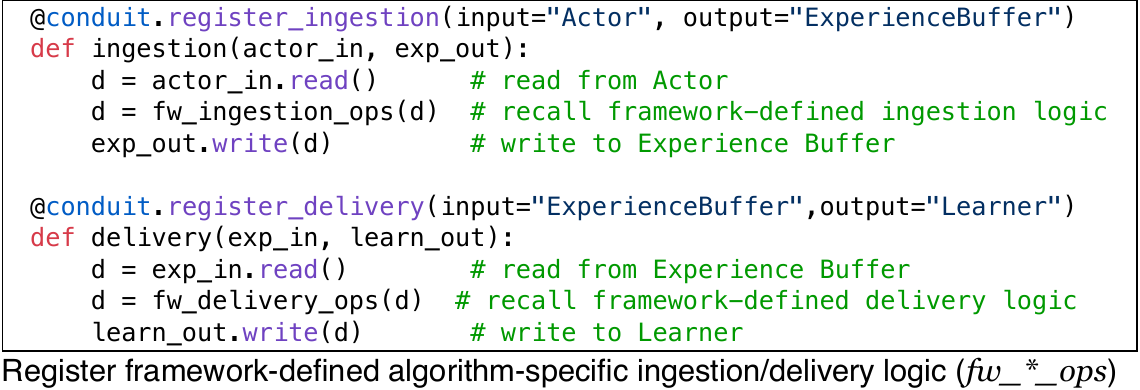}
\caption{\sys integration interface: frameworks register ingestion/delivery logic via buffer hooks; \sys controls placement and scheduling.}
	\label{fig:interface}
\end{figure}

\sys integrates with existing RL frameworks through a low-intrusion interception layer (\F\ref{fig:interface}).
Frameworks register their algorithm-specific ingestion and delivery logic via standard buffer hooks, while \sys owns execution timing, granularity, and capacity distribution.
This separation lets frameworks keep defining \emph{which} experiences are stored and consumed, while \sys controls \emph{where} they reside and \emph{when} experience-path handling executes, leaving framework logic unchanged.

\sys addresses the capacity demands and data movement costs of large-scale RL through \emph{capacity-constrained, bandwidth-aware placement}.
Using measured interconnect bandwidths and device-memory limits, \sys systematically distributes the Experience Buffer across CPU and GPU memory tiers.
This enables \sys to host experience datasets that exceed any single GPU's memory by distributing them across devices, while actively avoiding the slowest transfer edges on heterogeneous, non-uniform fabrics (\S\ref{sec:placement}).

Finally, \sys implements \emph{latency-aware scheduling} for experience-path handling.
It dynamically determines \emph{when} experience ingestion and delivery execute (execution timing) and \emph{how much} data they process per invocation (execution granularity).
This cost-based scheduling lets \sys overlap experience-path handling with actor rollout and learner update whenever semantics allow, amortizing overheads through batching and minimizing exposed experience-path latency on the critical path (\S\ref{sec:scheduling}).

%% file: sections/decoupling.tex
\section{Experience Data Plane}
\label{sec:decouple}

We formalize \sys's decoupling through the \emph{Experience Data Plane} (EDP), which breaks the tightly coupled actor\,$\rightarrow$\newline{}Experience Buffer\,$\rightarrow$\,learner dependency by turning experience ingestion, placement, and delivery into first-class asynchronous runtime operators rather than inline steps in framework execution loops.
This section focuses on the timing-related controls: ingestion, which governs how rollouts enter the buffer (\S\ref{subsec:actor_buffer}), and delivery, which governs how training-ready batches leave (\S\ref{subsec:learner_buffer}); placement (\emph{where} data resides) is deferred to \S\ref{sec:placement}.
This decoupling underpins \sys's capacity-constrained, bandwidth-aware placement (\S\ref{sec:placement}) and latency-aware scheduling (\S\ref{sec:scheduling}). \looseness=-1

\subsection{Decoupling Experience Ingestion}
\label{subsec:actor_buffer}

\noindent\textbf{Concept.}
Experience ingestion is the control point that separates actor execution from experience-path handling.
By making ingestion an explicit runtime operator, EDP lets actors push raw payloads asynchronously while ingestion pulls and processes data at its own pace.
Users only provide standard RL settings (e.g., rollout configuration and whether training is on/off-policy), and \sys derives valid experience-path control choices automatically.
We capture this knob using an \emph{ingestion granularity} $g_{\text{ing}}$, defined as how much rollout data one ingestion invocation processes, measured in multiples (or fractions) of one rollout unit.\footnote{A rollout unit is the RL algorithm-defined volume generated by one environment interaction phase, e.g., a rollout batch.}
 
\begin{figure*}[t!]
	\centering 
	\includegraphics[width=0.48\linewidth]{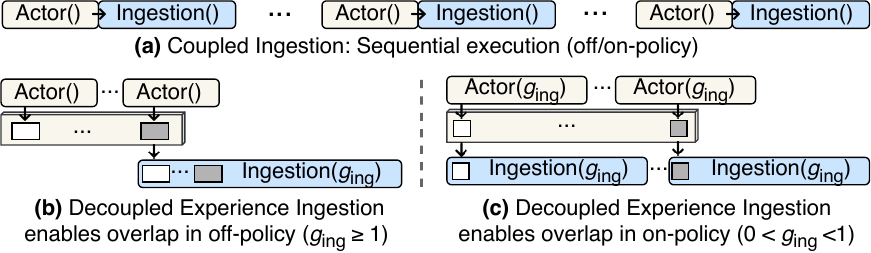}
	\hfill
	\includegraphics[width=.5\linewidth]{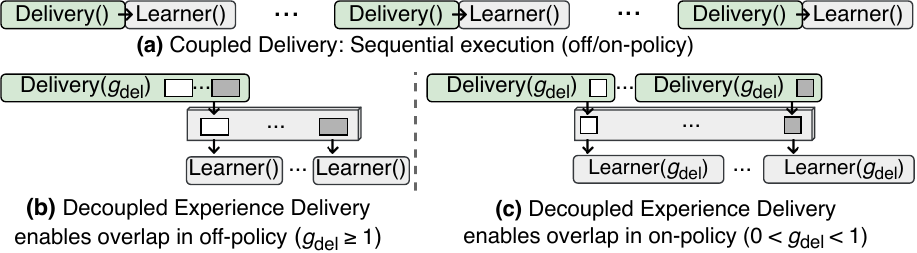}
	\caption{Decoupling ingestion (delivery) from actor (learner) execution.
	Actors push rollouts asynchronously, while the ingestion operator pulls data at configurable granularity, enabling batching and pipelining (left). The delivery operator prepares training-ready batches ahead of time, freeing learners from synchronous data-preparation waits (right).
	}\label{fig:decoupling}
\end{figure*}

\smallskip
\noindent\textbf{Enabling overlap and pipelining.}
Without decoupled ingestion%
, each rollout unit must be ingested immediately after it is generated, creating a strict one-to-one dependency between actor execution and experience handling.
With decoupled ingestion%
, actors push asynchronously, and $g_{\text{ing}}$ determines \emph{how} ingestion runs:
(i) when $g_{\text{ing}}\!\ge\!1$, ingestion can \emph{batch} multiple rollout units and process them together%
;
(ii) when $0\!<\!g_{\text{ing}}\!<\!1$, ingestion can \emph{pipeline} by processing a fraction of the current rollout while actor--environment interaction continues. %
This turns tight per-rollout dependencies into \emph{pipelineable constraints}, enabling overlap without changing RL semantics (left side of \F\ref{fig:decoupling}).

\smallskip
\noindent\textbf{Unifying off-policy and on-policy via $g_{\text{ing}}$.}
The feasible range of $g_{\text{ing}}$ encodes algorithm semantics (freshness vs.\ reuse) in a runtime boundary rather than hard-coding them into framework-specific actor logic.
\textit{Off-policy} training tolerates mild staleness, so ingestion may batch rollouts with
$g_{\text{ing}} \in \{1,\dots,g_{\text{ing}}^{\max}\}$, where $g_{\text{ing}}^{\max}$ caps staleness and memory overhead. 
\textit{On-policy} training requires fresh samples: if a rollout unit feeds $M$ mini-batch updates, ingestion can pipeline only at mini-batch granularity with
$g_{\text{ing}} \in \{\tfrac{1}{M},\tfrac{2}{M},\dots,1\}$. 

\subsection{Decoupling Experience Delivery}
\label{subsec:learner_buffer}
\noindent\textbf{Concept.}
Experience delivery separates experience-path handling from learner execution.
As an explicit runtime operator, it lets EDP prepare training-ready batches ahead of time while learners asynchronously pull staged batches.
Users specify only standard training configuration, and \sys derives valid control choices without manual tuning.
We define \emph{delivery granularity} $g_{\text{del}}$ as the amount of training data prepared per delivery invocation, measured in multiples or fractions of one training batch.

\noindent\textbf{Enabling overlap and pipelining.}
Without decoupled delivery, each learner update waits for the next batch, leaving little scheduling freedom.
With EDP boundaries, delivery runs proactively and stages batches, while $g_{\text{del}}$ sets the work per invocation:
when $g_{\text{del}}\!\ge\!1$, delivery can prefetch multiple batches, overlapping learner updates with delivery of upcoming batches;
when $0\!<\!g_{\text{del}}\!<\!1$, delivery feeds mini-batch chunks, enabling fine-grained intra-iteration pipelining (right side of \F\ref{fig:decoupling}).

\noindent\textbf{Unifying off-policy and on-policy via $g_{\text{del}}$.}
As with ingestion, the feasible range of $g_{\text{del}}$ captures semantic constraints at the EDP boundary.
For \textit{off-policy} training, delivery may prefetch multiple batches with
$g_{\text{del}}\in\{1,2,\dots,g_{\text{del}}^{\max}\}$, where $g_{\text{del}}^{\max}$ bounds staleness and memory overhead.
For \textit{on-policy} training, delivery preserves freshness within the current iteration and provides mini-batches sequentially, so $g_{\text{del}}=\tfrac{1}{M}$.

\subsection{Configuration and Correctness}
\label{subsec:config}

Users do \emph{not} tune buffer granularities directly.
Instead, \sys derives feasible settings from RL configurations that frameworks already expose: whether training is \emph{off-policy} or \emph{on-policy}, and (for on-policy) the mini-batch count $M$ per update.
These inputs define safe batching and mini-batch pipelining ranges that preserve replay and freshness semantics.
Within this safe space, \sys selects execution mode and effective granularity automatically; users may optionally cap batching to limit staleness or memory overhead, but sensible defaults typically suffice.
These same bounds also preserve correctness: EDP changes \emph{when} and \emph{where} experience-path handling runs, not \emph{what} data it produces, and overlap/granularity choices are drawn only from spaces that preserve on-policy freshness and off-policy replay semantics.
Formal invariants are summarized in Appendix~\ref{app:correctness}.

Overall, EDP turns framework execution loops into \emph{pipelineable constraints}, making experience-path operations independently schedulable while preserving RL semantics.

%% file: sections/placement.tex
\section{Capacity-Constrained, Bandwidth-Aware Experience Data Placement}
\label{sec:placement}

Building on the Experience Data Plane (\S\ref{sec:decouple}), \sys treats experience placement as an explicit optimization that reduces actor\,$\rightarrow$\,Experience Buffer\,$\rightarrow$\,learner overhead.
For each candidate placement, it (i) estimates transfer latency from measured effective bandwidths (\S\ref{subsec:transfer_model}) and (ii) selects the lowest-latency placement that remains feasible under memory constraints (\S\ref{subsec:placement_strategy}).
Since experience sizes can grow mid-run (\eg richer observations or longer sequences), an initial placement may become infeasible; \sys then performs (iii) online migration driven by a pre-computed placement map, avoiding runtime re-profiling (\S\ref{subsec:online_migration}).

\subsection{Transfer-Latency Model}
\label{subsec:transfer_model}

An Experience Buffer placement $p$ induces two critical-path transfers:
(\textbf{i}) moving new experiences from actors into the buffer (experience ingestion), and
(\textbf{ii}) moving sampled training data from the buffer to learner GPUs (experience delivery).
We model per-sample transfer latency as the sum of these two legs:
$T_{\text{transfer}}(p)=\frac{x}{BW(\text{actor}\rightarrow p)}+\frac{x}{BW(p\rightarrow \text{GPU})}$,
where $x$ is per-sample size and $BW(\cdot)$ is the \emph{effective bandwidth} of the path.
Effective bandwidth is obtained from microbenchmarks on the deployment stack and captures realized throughput under the hardware topology and runtime implementation~\cite{li2019evaluating,li2018tartan,wang2016interconnect}.

To account for locality (intra- vs.\ inter-node and GPU-rank dependence), $BW(\cdot)$ is looked up from measured bandwidth tables indexed by endpoint type and topology.
We consider four placement options: CPU pageable, CPU pinned, single GPU, and shared GPU. These cover the two factors that govern transfer cost on heterogeneous fabrics: memory tier (pageable/pinned CPU vs.\ GPU) and device distribution (single device vs.\ sharded across multiple devices). Together with the actor-side device type, each placement uniquely determines both legs, making $T_{\text{transfer}}(p)$ directly computable.

\tinyskip
\textbf{Example.} Consider a node with 8 GPUs wherein actors run on CPU, the Experience Buffer is sharded across GPUs, and learners run on GPUs.
Then $\text{actor}\,\rightarrow\, p$ uses CPU\,$\rightarrow$\,GPU bandwidth, while $p\,\rightarrow\,\text{GPU}$ uses GPU\,$\rightarrow$\,GPU bandwidth.
If a learner samples locally with probability $1/8$ (device-local copy) and remotely with probability $7/8$ (GPU\,$\rightarrow$\,GPU peer-to-peer), estimated from buffer sharding and sampling behavior, the expected bandwidth $BW(p\,\rightarrow\, \text{GPU})$ reflects this mixture and yields an accurate latency estimate.

\subsection{Placement Optimization}
\label{subsec:placement_strategy}

Given the transfer model above, \sys selects the placement that minimizes transfer latency subject to buffer capacity.
Let $C_{\text{buf}}$ be the required buffer capacity, $x$ the per-sample size, and $Mem(p)$ the profiled memory available under placement $p$.
Among candidate placements $\mathcal{P}$, \sys solves:
\begin{equation}
    \label{equ:placement_opt_clean}
    p^* \;=\; \arg\min_{p\in\mathcal{P}} \; T_{\text{transfer}}(p)
    \quad \text{s.t.}\quad C_{\text{buf}} \cdot x \le Mem(p).
\end{equation}

\noindent
\textbf{Configuration transparency.}
Users do not tune placement directly.
\sys derives actor/learner device types and $C_{\text{buf}}$ from the RL configuration, profiles $Mem(\cdot)$ and effective $BW(\cdot)$ at initialization, and solves Eq.~\eqref{equ:placement_opt_clean} automatically.
Users may optionally restrict candidates, but defaults typically suffice. %
Although Eq.~\eqref{equ:placement_opt_clean} defines the target optimum, exhaustive search is impractical: effective bandwidths vary with contention and concurrency, and device-subset candidates can grow as $\mathcal{O}(2^n)$ for $n$ devices.
\sys therefore uses an approximate, hardware-aware strategy (\A\ref{alg:placement}) that searches a small candidate set.

\mypar{Approximate solution}
\sys chooses placement at startup and during occasional reconfiguration, off the training critical path.
It constructs $\mathcal{P}$ using two heuristics.
(\textbf{i}) \emph{Cross-node balance:} in multi-node runs, \sys distributes capacity across nodes to avoid inter-node hotspots, since inter-node links are slower than intra-node GPU$\leftrightarrow$GPU fabrics.
(\textbf{ii}) \emph{Intra-node representatives:} within a node, \sys considers CPU pinned placement and shared GPU placements over GPU subsets of increasing size, preferring high-bandwidth ranks on non-uniform fabrics.
With $G\le8$ GPUs per node, enumerating these subsets remains efficient.
Given this reduced $\mathcal{P}$, \sys scans candidates (\A\ref{alg:placement}, lines~\ref{line:2}--\ref{line:5}) and selects the feasible placement with the smallest $T_{\text{transfer}}(p)$.

\mypar{Complexity}
Each $T_{\text{transfer}}(p)$ evaluation is constant-time, so selection costs $\mathcal{O}(|\mathcal{P}|)$.
In practice, $|\mathcal{P}|$ is small and bounded (\eg at most $2^G$ GPU subsets per node with $G\le8$), and placement runs outside the critical path.

\begin{figure*}[!ht]
	\centering
	\includegraphics[width=0.9\linewidth]{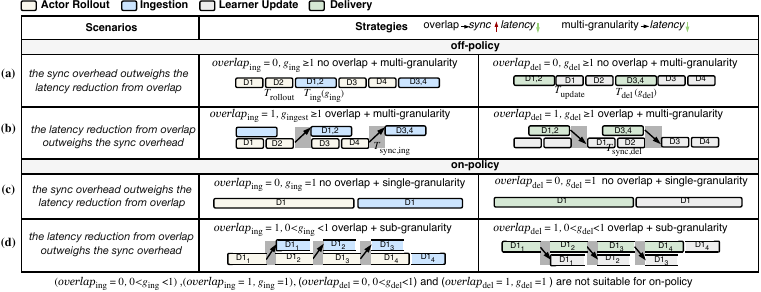}
	\caption{Ingestion\,/delivery-path modes: overlap $\times$ granularity across iterations (D1--D4) and mini-batches (D1$_1$--D1$_4$).}
	\label{fig:overlap-granularity}
\end{figure*}

\begin{algorithm}[t]
	\begin{footnotesize}
		\caption{\small Bandwidth-aware placement (linear scan): select the placement with minimum transfer latency.}
		\label{alg:placement}
		\KwIn{Actor-side device type, required buffer capacity $C_{\text{buf}}$, candidates $\mathcal{P}$, profiled $BW(\cdot)$ and $Mem(\cdot)$, per-sample size $x$. \looseness=-1}
		\KwOut{Selected placement $p^*$}
		$p^* \gets \texttt{None}$; $\;T_{\min}\gets +\infty$\;
		\ForEach{$p\in\mathcal{P}$}{ \label{line:2}
			\If{$C_{\text{buf}} \cdot x \le Mem(p)$}{ \label{line:3}
				$T \gets T_{\text{transfer}}(p)$\; \label{line:4}
				\textbf{if} $T < T_{\min}$ \textbf{then} $T_{\min}\gets T$, $p^*\gets p$\; \label{line:5}
			}
		}
		\Return $p^*$\;
	\end{footnotesize}
\end{algorithm}

\subsection{Online Placement Migration}
\label{subsec:online_migration}

The placement optimizer selects the initial placement, but in some RL workloads the per-sample size $x$ grows during training as observations, sequences, or auxiliary fields expand~\cite{sheng2025hybridflow,zhong2025optimizing}.
When growth makes the current placement infeasible, \sys switches via a pre-computed placement map to the next feasible low-latency placement, avoiding full online re-optimization.

At runtime, migration briefly pauses ingestion and delivery, transfers experience state using a pre-solved plan, and resumes both operators under the new placement.
Since the EDP decouples actor and learner execution, this adaptation does not stall either side of the training loop.
Appendix~\ref{app:migration} gives the migration algorithm and transfer-plan formulation.

%% file: sections/scheduling.tex
\section{Latency-Aware Scheduling of Experience Ingestion and Delivery}
\label{sec:scheduling}

With placement fixed (\S\ref{sec:placement}), \sys further reduces \emph{exposed} experience-path latency by scheduling \emph{experience ingestion} and \emph{delivery}.
Because the EDP (\S\ref{sec:decouple}) makes these operators independently runnable, \sys controls \emph{when} they run (\textit{timing}: sequential vs.\ overlapped) and \emph{how much} each invocation processes (\textit{granularity}: fractional vs.\ batched).
We outline these opportunities (\S\ref{subsec:pipeline}), model their latency impact (\S\ref{subsec:latency}), and formulate a cost-based optimizer for the latency-minimizing schedule (\S\ref{subsec:schedule}). \looseness=-1

\subsection{Scheduling Opportunities} 
\label{subsec:pipeline}

Each RL iteration traverses two EDP paths:
the \textbf{ingestion path} (actor rollout$\rightarrow$experience ingestion) and
the \textbf{delivery path} (experience delivery$\rightarrow$learner update).
The EDP makes ingestion and delivery \emph{schedulable} rather than inline with rollout/update, enabling cross-iteration batching (off-policy) or intra-iteration pipelining (on-policy) to reduce latency.

\noindent\textbf{Two semantic-free controls.}
\sys schedules both operators with two knobs:
(i) \emph{Timing}, controlled by $overlap_{\text{ing}}$ and $overlap_{\text{del}}\in\{0,1\}$, decides whether ingestion/delivery overlaps with rollout/update; and
(ii) \emph{Granularity}, controlled by $g_{\text{ing}}$ and $g_{\text{del}}$ (\S\ref{sec:decouple}), decides how much work each invocation performs.
Feasible granularities are constrained by $\mathcal{G}_{\text{ing}}$ and $\mathcal{G}_{\text{del}}$ derived from standard RL configuration (\S\ref{subsec:schedule}), keeping scheduling algorithm-agnostic.

Fig.~\ref{fig:overlap-granularity} instantiates these knobs as execution modes.
Non-overlapped timing corresponds to (a) and (c), while overlap corresponds to (b) and (d).
Along granularity, $g>1$ batches units in (a) and (b), $g=1$ runs single-unit in (c), and $0<g<1$ pipelines mini-batches in (d).
Together, these knobs cover the common RL settings---off-policy sequential/overlapped and on-policy sequential/pipelined---under a unified scheduling interface.

\subsection{Iteration-Latency Model}
\label{subsec:latency}

We model one RL iteration as $T_{\text{iter}} = T_{\text{ing-path}} + T_{\text{del-path}}$,
where $T_{\text{ing-path}}$ (resp.\ $T_{\text{del-path}}$) is the exposed latency of the ingestion (resp.\ delivery) path.
Let $T_{\text{rollout}}$ and $T_{\text{update}}$ denote the execution time of one actor rollout and one learner update.
Let $T_{\text{ing}}(g)$ and $T_{\text{del}}(g)$ denote the wall-clock time of one ingestion/delivery invocation that processes $g$ rollout units/training batches (including transfer and processing under placement $p^*$ from \S\ref{sec:placement}).

\tinyskip
\noindent\textbf{Granularity and timing.}
Processing more data per invocation amortizes fixed per-invocation overheads, so we use per-unit costs $\bar{T}_{\text{ing}}(g)=T_{\text{ing}}(g)/g$ and $\bar{T}_{\text{del}}(g)=T_{\text{del}}(g)/g$.
Overlap hides part of ingestion/delivery behind rollout/update, but adds synchronization overhead.

Given a scheduling mode, the exposed ingestion-path time is
$T_{\text{ing-path}} = T_{\text{rollout}} + \bar{T}_{\text{ing}}(g_{\text{ing}})
+ overlap_{\text{ing}}\!\cdot\!\bigl(T_{\text{sync,ing}} - hide_{\text{ing}}\bigr)$,
and the exposed delivery-path time is
$T_{\text{del-path}} =
T_{\text{update}}
+\bar{T}_{\text{del}}(g_{\text{del}})
+ overlap_{\text{del}}\!\cdot\!\bigl(T_{\text{sync,del}} - hide_{\text{del}}\bigr)$.
The hidden terms $hide_{\text{ing}}$ and $hide_{\text{del}}$ capture the overlapable portion of operator latency under off-policy batching and on-policy pipelining; Appendix~\ref{app:latency-model} details these cases.
This compact model matches \F\ref{fig:overlap-granularity}: larger $g$ lowers amortized cost, while overlap hides latency at the cost of synchronization.
\subsection{Cost-Based Scheduling}
\label{subsec:schedule}
\sys minimizes iteration latency by jointly choosing overlap and granularity for both operators:
{\small
\begin{equation}
	\label{eq:sched_opt}  
	\begin{aligned}
		\min \;\; & T_{\text{ing-path}}(overlap_{\text{ing}}, g_{\text{ing}})
		+ T_{\text{del-path}}(overlap_{\text{del}}, g_{\text{del}}) \\
		\text{s.t.}\;\; &
		overlap_{\text{ing}}, overlap_{\text{del}} \in \{0,1\},
	    g_{\text{ing}} \in \mathcal{G}_{\text{ing}},\;
		g_{\text{del}} \in \mathcal{G}_{\text{del}} .
	\end{aligned}
\end{equation}
} 
\noindent\textbf{Configuration transparency.}
Users do not tune overlap flags or granularities directly.
\sys derives $\mathcal{G}_{\text{ing}}$ and $\mathcal{G}_{\text{del}}$ from RL configuration---integer granularities for off-policy replay and fractional mini-batch steps for on-policy training---optionally respecting user caps.
It then profiles $T_{\text{rollout}}$, $T_{\text{update}}$, and operator costs once, enumerates the resulting finite candidate set, and selects the lowest-latency mode under Eq.~\eqref{eq:sched_opt}.
\S\ref{subsec:exp_auto} shows it automatically selects the best configuration.

\tinyskip
\noindent\textbf{Complexity.}
\sys solves Eq.~\eqref{eq:sched_opt} in constant time, $\mathcal{O}(1)$.
It enumerates all $4\cdot|\mathcal{G}_{\text{ing}}|\cdot|\mathcal{G}_{\text{del}}|$ candidates (two binary overlap flags), scoring each in constant time from profiled costs; this count is fixed by the RL configuration and independent of workload scale, so the search runs once at initialization (and only re-runs if workload or resources change).

%% file: sections/evaluation.tex
\section{Evaluation}
\label{sec:results}

We evaluate \sys along the three dimensions of the Experience Data Plane introduced in \S\ref{sec:motivation}: (i) \emph{decoupling}, (ii) \emph{capacity-constrained, bandwidth-aware placement}, and (iii) \emph{latency-aware scheduling}.
After outlining the experimental setup (\S\ref{subsec:exp_setting}), the evaluation answers the following questions:

\noindent
(1) What is \sys's overall performance, and how does it reduce exposed experience-path latency in state-of-the-art distributed RL frameworks? (\S\ref{subsec:exp_overall})

\noindent
(2) How well does \sys's capacity-constrained, bandwidth-aware placement adapt to workloads under heterogeneous interconnect and device-memory limits? (\S\ref{subsec:exp_placement})

\noindent
(3) How effectively does \sys's latency-aware scheduling reduce exposed experience-path latency? (\S\ref{subsec:exp_scheduling})

\noindent
(4) How does \sys's auto-optimizer compare against tuned manual baselines? (\S\ref{subsec:exp_auto})

\noindent
(5) What is the scalability performance of \sys? (\S\ref{subsec:exp_scalability})

\noindent
(6) Does \sys preserve RL convergence? (\S\ref{subsec:exp_convergence})

\subsection{Experimental Setup}
\label{subsec:exp_setting}
Our experiments use the following setup.

\mypar{Testbeds} We conduct experiments on two clusters: \textbf{(1)} an NVIDIA A100 cluster---two nodes, each equipped with eight NVIDIA A100 80\,GB GPUs, two Intel Xeon Platinum 8342 CPUs (48 cores in total), and 2\,TB RAM, running Red Hat Enterprise 9.5; and \textbf{(2)} an AMD MI250X supercomputer---each node containing eight AMD EPYC 7A53 CPUs (512 cores total) and eight AMD MI250X GPUs (64\,GB per GPU die), running SUSE Linux Enterprise Server 15 SP5.
We use the A100 cluster for overall efficiency (\S\ref{subsec:exp_overall}), placement (\S\ref{subsec:exp_placement}), scheduling (\S\ref{subsec:exp_scheduling}), joint adaptive control (\S\ref{subsec:exp_auto}), and convergence (\S\ref{subsec:exp_convergence}).
We use the MI250X supercomputer for scalability experiments (\S\ref{subsec:exp_scalability}) to demonstrate hardware generality and leverage its larger GPU pool (up to 1{,}024 GPUs).

\tinyskip
\mypar{Baselines} We compare \sys against the following three baselines:
\texttt{RLlib}~\cite{liang2021rllib}, an open-source RL framework whose experience buffer is embedded in actor/learner control flow with default CPU placement and request-driven sampling;
\noindent
\texttt{Gear}~\cite{wang2023gear}, an RL buffer service that stores trajectories in pinned host memory and uses GPU-driven zero-copy DMA/RDMA to accelerate transfers; and
\texttt{Reverb}~\cite{cassirer2021reverb}, a distributed buffer service that hosts buffers in CPU memory for scalable ingestion.
We also compare against two other RL frameworks, \texttt{Verl}~\cite{sheng2025hybridflow} (LLM post-training) and \texttt{SRL}~\cite{mei2023srl}, in Appendices~\ref{app:llm-post-training} and~\ref{app:srl}.

\tinyskip
\mypar{RL algorithms}
We evaluate three representative RL algorithms: Deep Q-Network (DQN)~\cite{mnih2013playing} and soft actor-critic (SAC)~\cite{haarnoja2018soft} as off-policy workloads, and proximal policy optimization (PPO)~\cite{schulman2017proximal} as an on-policy workload.
Appendix~\ref{app:llm-post-training} additionally studies Group Relative Policy Optimization (GRPO)~\cite{shao2024deepseekmath} in an LLM post-training pipeline.

\mypar{Environments}
We evaluate five environments that span different kinds of experience-path pressure: MountainCar~\cite{classic}, a classic control task with small fixed-size observations; Multi-agent CartPole~\cite{terry2021pettingzoo}, a cooperative multi-agent task that multiplies per-step experience volume; Meta-World~\cite{yu2020meta} ($\sim$48\,KB per sample), a 50-task robotic manipulation benchmark with visual observations; Open X-Em\-bod\-i\-ment~\cite{o2024open} (OXE, $\sim$192\,KB per sample), a large-scale multi-embodiment robotics dataset aggregating 60+ real-robot sources; and a configurable synthetic stress test following~\cite{cassirer2021reverb}, whose transition dimension varies from 128 to 20{,}000 for the placement, scheduling, and adaptive-control studies.

\mypar{Metrics}
We use \textit{end-to-end iteration latency}---including actor rollout, experience-path handling, and learner update---as the primary metric.
We also measure the \textit{exposed experience-path latency}, which captures the portion of ingestion and delivery work that is not hidden by overlap with rollout or learner update within each iteration.

\begin{figure}[t]
    \centering
    \includegraphics[width=\linewidth]{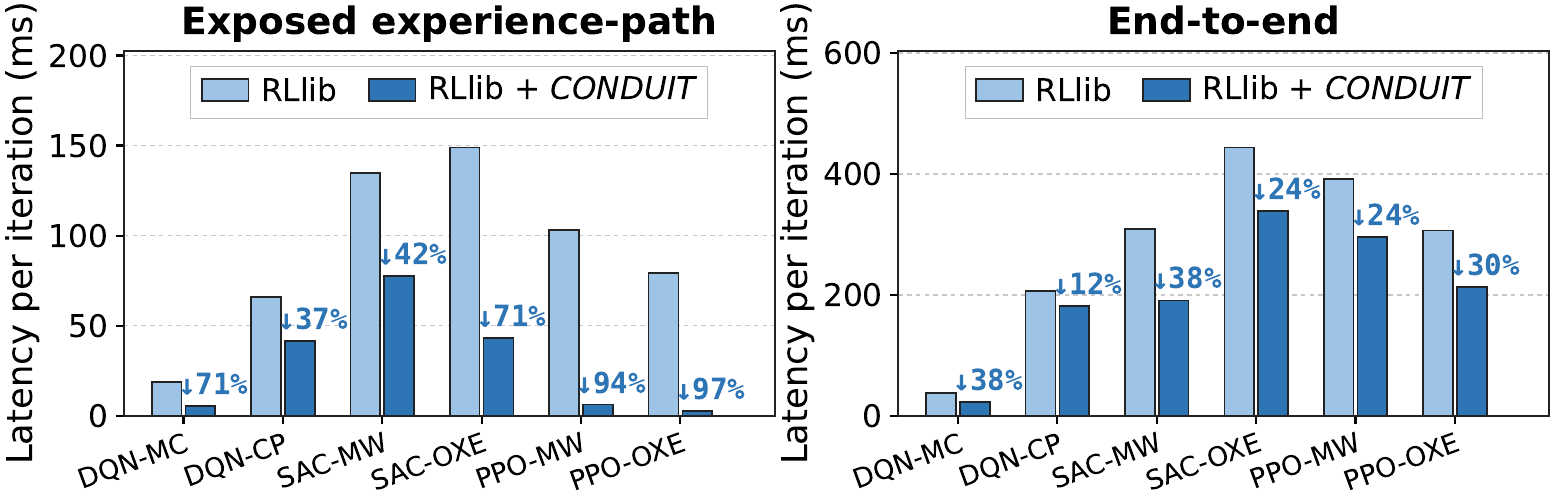}
    \caption{\sys's per-iteration latency reduction over \texttt{RLlib}: exposed experience-path (left), end-to-end (right). Workloads: \mbox{MC}=MountainCar, \mbox{CP}=MA-CartPole, \mbox{MW}=Meta-World, \mbox{OXE}=Open X-Embodiment}
    \label{fig:latency_all}
\end{figure}

\subsection{Overall Effectiveness and Efficiency}
\label{subsec:exp_overall}

We first evaluate \sys's end-to-end benefit by integrating it into \texttt{RLlib}, a production-grade distributed RL framework: \sys drop-in replaces RLlib's built-in experience buffer without modifying actor/learner control flow. This directly exercises the \emph{decoupling} dimension, validating that the Experience Data Plane is structurally independent of framework execution logic. We report both end-to-end iteration latency and exposed experience-path latency, and omit \texttt{Gear} and \texttt{Reverb}, which are standalone buffer services without compatible drop-in hooks. \looseness=-1

We test \sys across three algorithms (DQN, SAC, PPO) and four environments: DQN on \textit{MountainCar} and \textit{Multi-agent CartPole}, and SAC and PPO on \textit{Meta-World} and OXE (large visual observations that stress both capacity and bandwidth), using eight actors and eight learners.
 
\F\ref{fig:latency_all} reports exposed experience-path latency (left) and end-to-end iteration latency (right) for each (algorithm, environment) pair.
Across all workloads, integrating \sys with \texttt{RLlib} reduces exposed experience-path latency---by up to 71\% for DQN, 71\% for SAC, and 97\% for PPO.
These reductions translate into end-to-end improvements of up to 38\% (DQN), 38\% (SAC), and 30\% (PPO).
Two EDP levers drive these reductions: scheduling hides experience-path handling behind rollout and update, while bandwidth-aware placement distributes the large buffer across devices to keep transfer costs low.
For off-policy DQN and SAC the savings reach end-to-end latency, as replay adds cross-iteration overlap. For on-policy PPO, \sys cuts experience-path latency sharply, but end-to-end gains remain bounded since freshness limits batching and the residual rollout/update time dominates the critical path. \looseness=-1 

\mypar{Insights}
\sys's end-to-end benefit scales with the experience path's share of the iteration: the larger that share, the greater the gain from EDP's placement and scheduling optimizations.

\subsection{Bandwidth-Aware Placement}
\label{subsec:exp_placement}

Using the \textit{Synthetic Environment}, we sweep transition dimension from $128$ to $20{,}000$ with buffer capacity $10^6$ and batch size $512$.
This sweep covers both axes of the \emph{placement} dimension: for small transitions, the buffer fits within one GPU and placement simply optimizes bandwidth (\emph{bandwidth-dominated regime}); as transitions grow, the buffer exceeds per-GPU memory and capacity becomes the constraint---only placements with sufficient aggregate memory remain viable (\emph{capacity-dominated regime}).

\begin{figure}[t]
    \centering
    \includegraphics[width=0.85\linewidth]{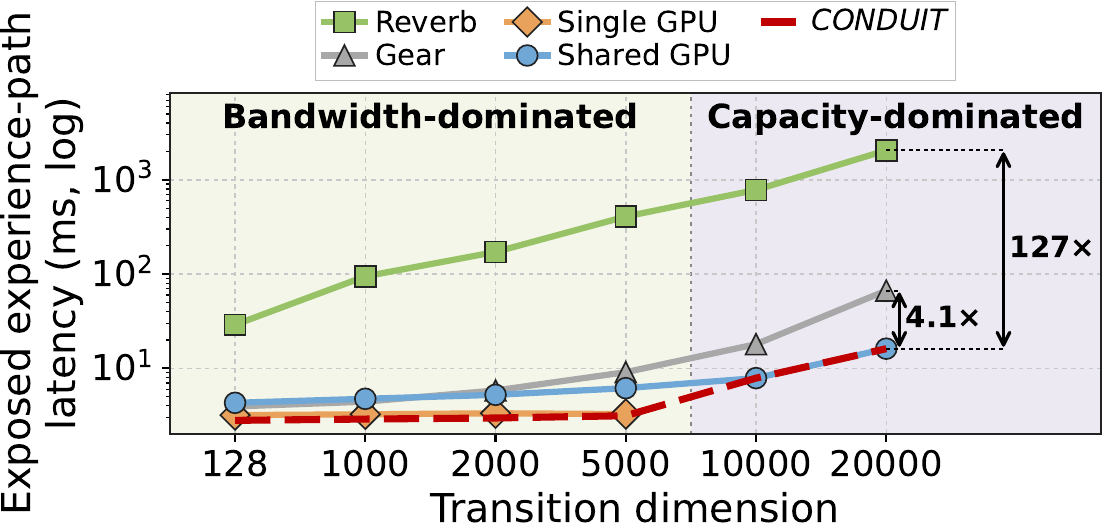}
    \caption{Placement ablation: \sys vs.\ four placements (\texttt{Reverb}=CPU pageable, \texttt{Gear}=CPU pinned with zero-copy access, single GPU, shared GPU). Missing points denote OOM.}
    \label{fig:placement_ablation}
\end{figure}

\begin{figure}[t]
    \centering
    \includegraphics[width=0.85\linewidth]{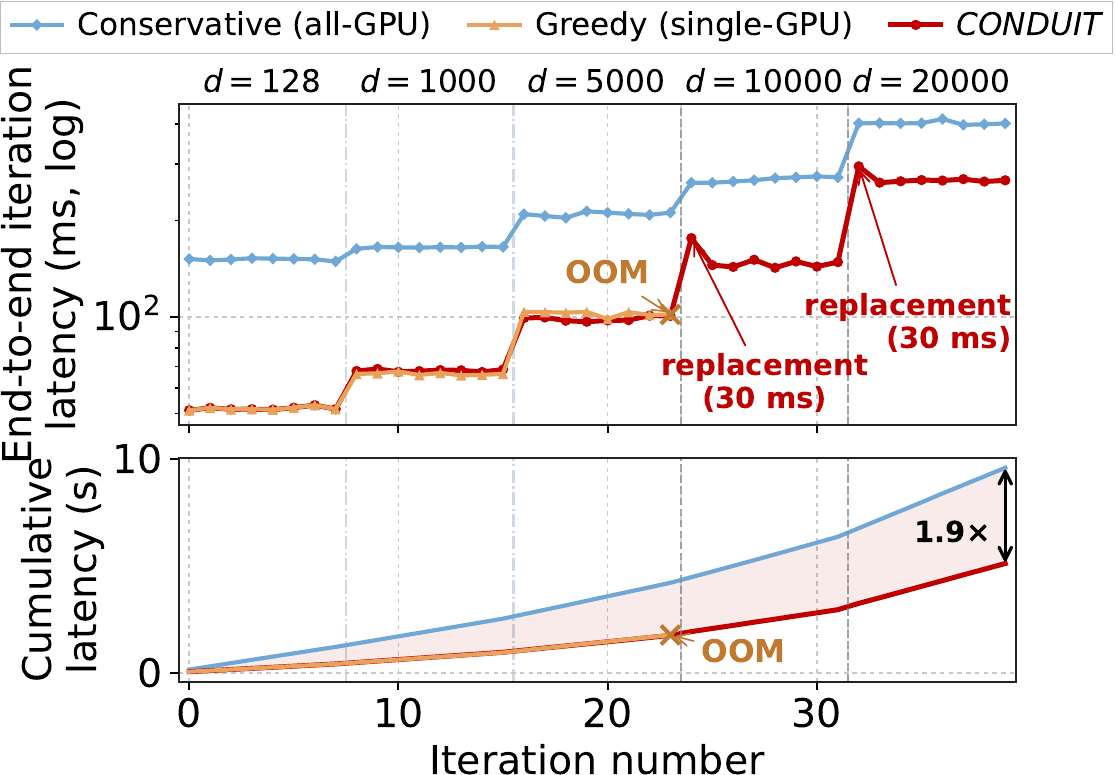}
    \caption{Online migration as transition dimension $d$ grows, vs.\ \emph{Greedy} (single GPU) and \emph{Conservative} (all-GPU) baselines. Top: iteration latency (log scale); bottom: cumulative latency.}
    \label{fig:migration}
    
\end{figure} 

\mypar{\sys vs.\ state-of-the-art buffer services}
\F\ref{fig:placement_ablation} compares \sys against the four placements that \texttt{Reverb} (CPU pageable) and \texttt{Gear} (CPU pinned, with GPU-driven zero-copy access) instantiate, alongside single-GPU and shared-GPU placement.
In the bandwidth-dominated regime ($128$--$5{,}000$), all placements are feasible; single-GPU placement is fastest by eliminating CPU--GPU transfers, so \sys adopts it, reducing latency by up to $130\times$ over \texttt{Reverb} and $2.9\times$ over \texttt{Gear} at $d{=}5{,}000$.
Beyond $5{,}000$, single GPU exceeds per-GPU memory and \sys switches to shared GPU, which dominates via fast GPU--GPU links ($\sim$10$\times$ faster than CPU--GPU), 
leaving \sys $127\times$ faster than \texttt{Reverb} and $4.1\times$ faster than \texttt{Gear} at $d{=}20{,}000$. These ratios primarily reflect memory-tier placement: both baselines keep experience CPU-resident, and the gap quantifies what bandwidth-aware, GPU-resident placement buys.

\mypar{Online migration}
\F\ref{fig:migration} traces this adaptive process over time: as transition dimension grows from $128$ to $20{,}000$, \sys automatically migrates from single GPU to shared GPU when the capacity boundary is crossed.
We compare against two baselines that bracket the trade-off.
\emph{Greedy} selects the lowest-latency placement for the initial workload (single GPU per learner); it is fastest while it fits ($52$--$103$\,ms) but exhausts GPU memory at $d{\ge}10{,}000$.
\emph{Conservative} shards the buffer across all $8$ GPUs from the start to guarantee feasibility at peak size; it stays feasible throughout but pays multi-GPU coordination overhead at every dimension ($151$--$403$\,ms).

\sys combines the strengths of both: it matches \emph{Greedy} while feasible, then absorbs two brief $\sim$30\,ms replacement spikes (at $d{=}10{,}000$ and $d{=}20{,}000$) and holds iteration latency at $146$ and $266$\,ms in the two largest segments, $34$--$46\%$ below \emph{Conservative} once \emph{Greedy} has already failed.
Each replacement reallocates the buffer under the new shard layout, reusing freed HBM from the caching allocator so the spike stays in the tens of milliseconds.
Over the full $40$-iteration trace, \sys accumulates $1.9\times$ less cumulative latency than \emph{Conservative}, while \emph{Greedy} cannot complete the run.

\mypar{Insights}
Capacity and bandwidth jointly determine placement: \sys's cost model selects the lowest-latency feasible placement at each operating point, and the same model drives online migration when workload statistics shift, keeping placement aligned throughout training. \looseness=-1 

\begin{figure*}[t]
    \centering

    \begin{minipage}[b]{0.32\linewidth}\centering
        \includegraphics[width=\linewidth]{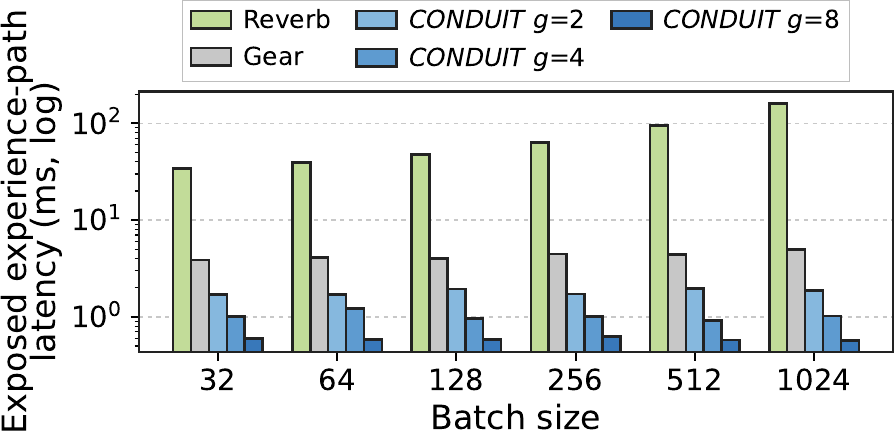}%
    \end{minipage}%
    \hfill%
    \begin{minipage}[b]{0.32\linewidth}\centering
        \includegraphics[width=\linewidth]{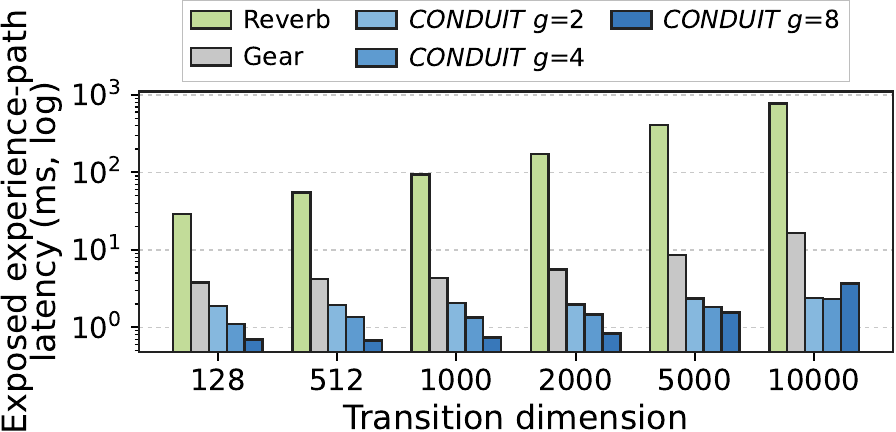}%
    \end{minipage}%
    \hfill%
    \begin{minipage}[b]{0.32\linewidth}\centering
        \includegraphics[width=\linewidth]{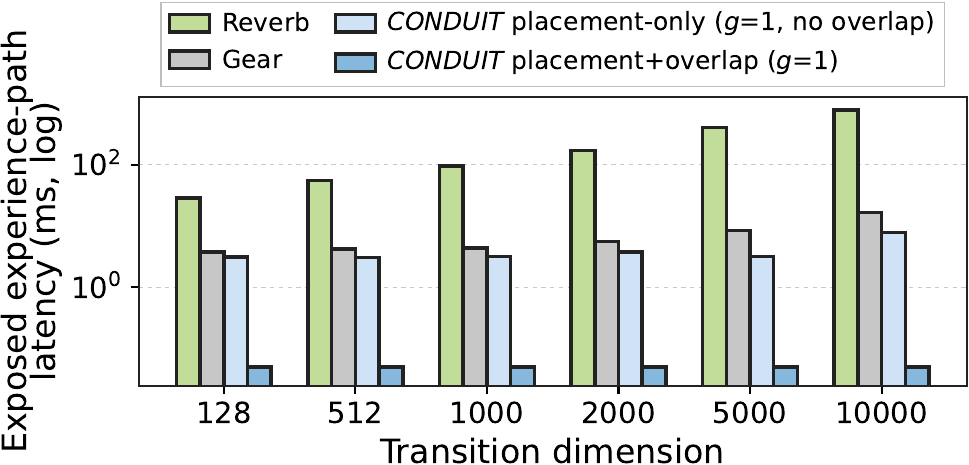}%
    \end{minipage}%

    \caption{Impact of granularity $g$ under varying batch sizes (left) and transition dimensions (middle) for $g$=$g_{\text{ing}}$=$g_{\text{del}}$; the right panel shows the impact of execution timing \text{\footnotesize{$overlap$}} under varying transition dimensions.}
    \label{fig:latency-aware}
\end{figure*}

\subsection{Latency-Aware Scheduling}
\label{subsec:exp_scheduling}

This section isolates the benefit of \sys's \emph{latency-aware scheduling}.
We study how the two control knobs---execution \emph{granularity} ($g$) and execution \emph{timing} ($overlap$)---affect exposed experience-path latency under increasing workload pressure, and compare against \texttt{Reverb} and \texttt{Gear} (\F\ref{fig:latency-aware}). \looseness=-1

\mypar{Execution granularity}
Using the \textit{Synthetic Environment}, we fix timing to sequential execution ($overlap=0$) and scale workload along two dimensions:
(i) batch size $32$--$1{,}024$ at fixed transition dimension $1{,}000$, and
(ii) transition dimension $128$--$10{,}000$ at fixed batch size $512$.

The left and middle panels of \F\ref{fig:latency-aware} report \sys's exposed experience-path latency at $g\in\{2,4,8\}$ against both baselines.
Larger $g$ amortizes per-invocation overheads (kernel launch, inter-process communication, metadata, inter-rank synchronization) and enables batched GPU throughput (\S\ref{subsec:latency}): on the batch sweep, exposed latency drops from $1.87$\,ms ($g{=}2$) to $0.57$\,ms ($g{=}8$) at batch $1{,}024$.
The gain then reverses on harder workloads: at dimension $10{,}000$ the optimum shifts back to $g{=}4$ ($2.33$\,ms) and $g{=}8$ regresses to $3.67$\,ms, as the per-invocation chunk grows large enough that managing it outweighs the amortization benefit.

\mypar{Execution timing} 
We next isolate timing by varying transition dimension from $128$ to $10{,}000$ and comparing four configurations:
\texttt{Reverb}, \texttt{Gear}, \sys placement-only ($g{=}1$, $overlap{=}0$), and \sys with overlap ($g{=}1$, $overlap{=}1$).
The right panel of \F\ref{fig:latency-aware} shows that \texttt{Reverb} and \texttt{Gear} (no overlap) suffer $27\times$ and $4.4\times$ growth in experience-path latency as transition dimension increases.
By contrast, \sys's placement-only mode grows only $2.5\times$ ($3.12$ to $7.83$\,ms) by holding experience handling in fast memory; adding overlap then hides most of this residual behind rollout and update, further reducing the exposed latency. 

\mypar{Insights}
\sys's scheduling wins by combining overlap and granularity: overlap hides ingestion/delivery behind rollout and update, while $g$ amortizes per-invocation overheads.

\subsection{Joint Adaptive Control: Auto vs Manual}
\label{subsec:exp_auto}

\begin{figure}[t]
    \centering
    \includegraphics[width=0.85\linewidth]{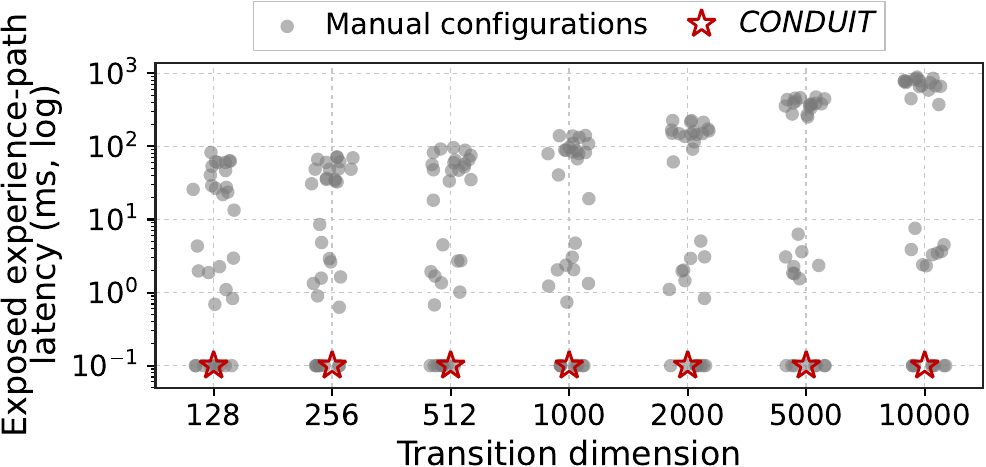}
    \caption{Auto vs.\ manual over the joint placement $\times$ overlap $\times$ granularity space. Gray: $32$ manual configs; red star: \sys's auto pick.}
    \label{fig:auto_vs_manual}

\end{figure}

This subsection evaluates whether \sys's auto-optimizer correctly navigates the \emph{joint} placement$\times$scheduling design space.
Using the \textit{Synthetic Environment}, we sweep $32$ manual configurations ($4$ placements $\times$ $2$ overlap settings $\times$ $4$ granularities) across transition dimensions $128$ to $10{,}000$, and compare them against \sys's automatic choice.
\F\ref{fig:auto_vs_manual} shows that the cloud span widens from $82.9$\,ms at dimension$=128$ to $888.5$\,ms at dimension$=10{,}000$, and that the lowest manual point is reached by \emph{different} configurations across dimensions: single GPU with overlap and $g{=}8$ at dimension$=128$, and single GPU with overlap and $g{=}2$ at dimension$=10,000$.
\sys's automatic choice sits on the lower envelope 
because the profile-driven cost model ranks candidates correctly at every operating point. 
Appendix~\ref{app:costmodel} give further details of the cost model's accuracy.

\mypar{Insights}
\sys's auto-optimizer matches the best manual point on every workload without re-tuning, while any fixed configuration must compromise across workloads.
 
\subsection{Scalability Analysis} 
\label{subsec:exp_scalability}

\begin{figure}[t]
	\centering
	\begin{subfigure}[t]{0.95\linewidth}
		\centering
		\includegraphics[width=0.85\linewidth]{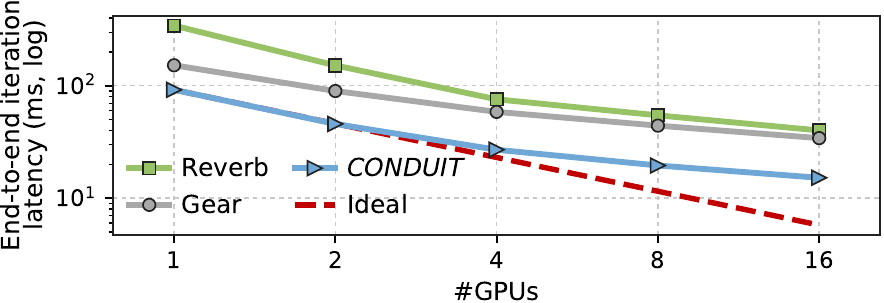}
		\caption{A100 cluster: 1--16 GPUs at batch size 2{,}048.}
		\label{fig:scalability_a100}
	\end{subfigure}
	
	\begin{subfigure}[t]{0.95\linewidth}
		\centering
		\includegraphics[width=0.85\linewidth]{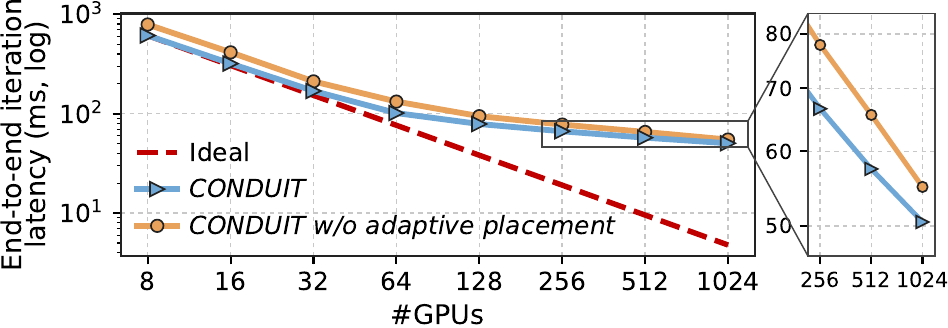}
		\caption{MI250X supercomputer: 8--1{,}024 GPUs at batch size 32{,}768.}
		\label{fig:scalability_mi250x}
	\end{subfigure}
	\caption{Strong scaling of \sys with GPU count on an NVIDIA A100 cluster and an AMD MI250X supercomputer (end-to-end iteration time, log scale).}
	\label{fig:scalability}
\end{figure}

We evaluate \sys's strong-scaling behavior on two platforms (\F\ref{fig:scalability}).
On the A100 cluster, we fix the total rollout/training batch size to 2{,}048 and scale from 1 to 16 A100 GPUs, comparing against two buffer-service baselines, \texttt{Gear} and \texttt{Reverb}.
On the MI250X supercomputer, we scale \sys from 8 to 1{,}024 MI250X GPUs under a fixed workload of 32{,}768 and compare against \sys-\texttt{\small{w/o-adaptive\-placement}}, which disables adaptive placement. \looseness=-1

\F\ref{fig:scalability_a100} shows that \sys scales more efficiently than both buffer-service baselines.
From 1 to 16 GPUs, \sys reduces end-to-end iteration latency from 95\,ms to 15\,ms (84\% reduction), remaining consistently below both \texttt{Gear} and \texttt{Reverb} at every scale.
At 16 GPUs, \sys achieves 57\% lower latency than \texttt{Gear} (15 vs.\ 35\,ms) and 65\% lower than \texttt{Reverb} (15 vs.\ 43\,ms).
All systems deviate from ideal linear scaling at larger GPU counts due to synchronization overhead, but \sys maintains the smallest gap.

\F\ref{fig:scalability_mi250x} shows that \sys scales into the supercomputer regime.
From 8 to 1{,}024 GPUs, \sys cuts end-to-end iteration latency from 614\,ms to 50\,ms (92\%), and beats \sys-\texttt{w/o-adaptive-placement} by 9--21\% at all counts (e.g., 110 vs.\ 140\,ms at 64 GPUs, 50 vs.\ 55\,ms at 1{,}024).
This gap reflects adaptive placement on the MI250X's non-uniform intra-node fabric (links of roughly 100--400\,GB/s): \sys prefers higher-bandwidth GPU ranks (0/2/4/6 in \F\ref{fig:fabric_topo}) over slower ones (1/3/5/7).
We omit \texttt{Gear} and \texttt{Reverb} here, as they lack AMD support~\cite{reverb-amd-issue,gear-github}; this experiment thus shows hardware generality and adaptive placement at scale rather than a head-to-head comparison.
 
\mypar{Insights}
\sys scales across both NVIDIA and AMD platforms, staying faster than baselines at every scale.
It tracks the ideal trend at small scales and keeps a smaller gap at large scales by reducing exposed experience-path latency on the critical path.
On the MI250X supercomputer, adaptive placement further improves scaling by avoiding slow ranks under non-uniform intra-node links.

\subsection{Convergence and Correctness Analysis}
\label{subsec:exp_convergence}
\begin{figure}[t]
    \centering
    \begin{subfigure}[t]{0.47\linewidth}
        \centering
        \includegraphics[width=\linewidth]{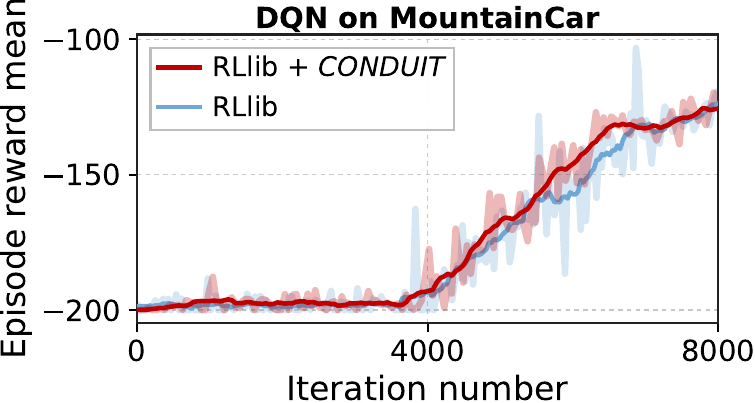}
        \caption{Reward vs.\ training iteration.}
        \label{fig:convergence_step}
    \end{subfigure}
    \hfill
    \begin{subfigure}[t]{0.47\linewidth}
        \centering
        \includegraphics[width=\linewidth]{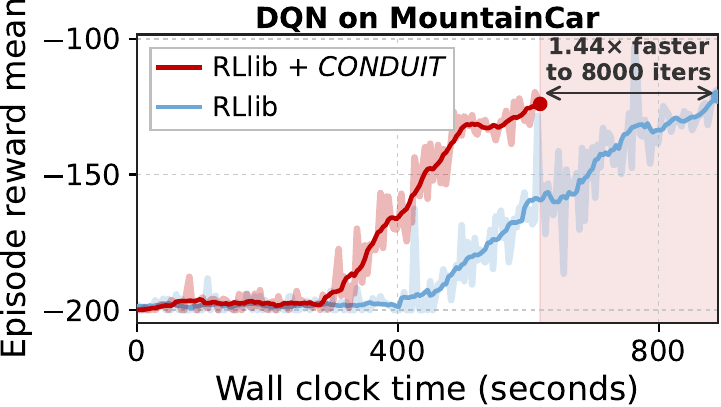}
        \caption{Reward vs.\ wall-clock time.}
        \label{fig:convergence_time}
    \end{subfigure}

    \vspace{0.6em}
    \begin{subfigure}[t]{0.47\linewidth}
        \centering
        \includegraphics[width=\linewidth]{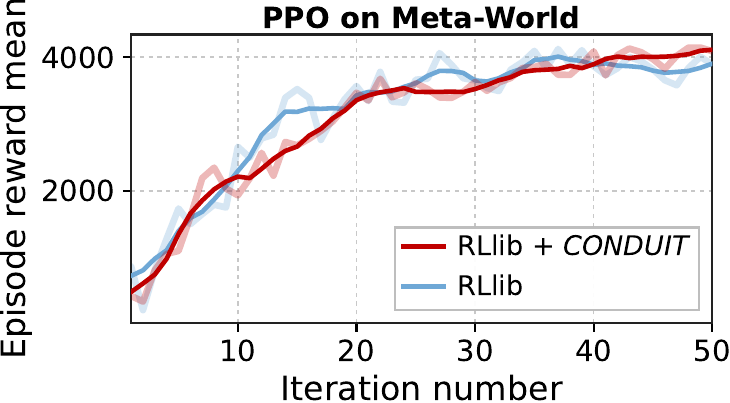}
        \caption{Reward vs.\ training iteration.}
        \label{fig:convergence_step_ppo}
    \end{subfigure}
    \hfill
    \begin{subfigure}[t]{0.47\linewidth}
        \centering
        \includegraphics[width=\linewidth]{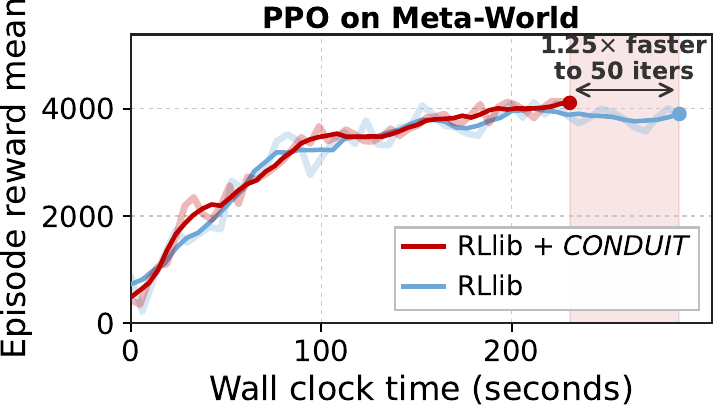}
        \caption{Reward vs.\ wall-clock time.}
        \label{fig:convergence_time_ppo}
    \end{subfigure}
    \caption{Convergence analysis for \sys on RLlib.}
    \label{fig:convergence}
    \vspace{-2.5em}
\end{figure}

We evaluate (\F\ref{fig:convergence}) whether \sys changes RL training behavior by comparing the learning curves of \texttt{RLlib} with and without \sys on DQN/\textit{MountainCar} (off-policy) and PPO/\textit{Meta-World} (on-policy), each with eight actors and eight learners.
The left column reports episode reward versus training iteration; \sys closely matches the baseline trend on both tasks, indicating no observable impact on convergence.
The right column reports the same metric versus wall-clock time; \sys reaches comparable reward levels sooner.
This time-to-quality gain comes from reducing exposed experience-path latency, which shortens iteration time without changing algorithm semantics.
Together with the invariant-preserving EDP boundary in \S\ref{sec:decouple}, this result supports semantics preservation across the evaluated on-policy and off-policy settings.

%% file: sections/related.tex
\section{Related Work}
\label{sec:related} 

\mypar{Replay buffers and distributed RL systems}
Large RL systems decouple actors and learners via replay buffers and actor/learner parallelism---Ape-X~\cite{horgan2018distributed}, importance-weighted actor-learner architectures (IMPALA)~\cite{espeholt2018impala}, and SEED RL~\cite{espeholt2019seed}.
Other work scales multi-GPU distributed reinforcement learning (DRL) via finer-grained GPU sharing~\cite{wang2025gmidrl}, and reinforcement learning from human feedback (RLHF)/large-model stacks optimize model execution and orchestration (\texttt{RLlib}~\cite{liang2021rllib}, \texttt{Verl}~\cite{sheng2025hybridflow}, PUZZLE~\cite{lei2024puzzle}), but these efforts target compute rather than the experience path.
Replay-buffer services (Reverb~\cite{cassirer2021reverb}) provide APIs and sampling but treat the buffer as passive storage driven by actor/learner loops.
\sys is complementary: it treats the experience path itself as the optimization target, cutting \emph{exposed} latency without changing algorithm logic.
 
\mypar{Decoupling and pipelining for RL training}
Asynchronous actor--learner execution improves throughput (A3C~\cite{mnih2016asynchronous}, IMPALA~\cite{espeholt2018impala}), but experience-path handling often runs inline or as fixed background behavior, limiting control over \emph{when} it runs and \emph{how much} it processes. \sys exposes this control explicitly through EDP. 

\mypar{Data movement and placement in heterogeneous training stacks}
Prior work reduces training overhead by optimizing placement and movement across heterogeneous memory and network fabrics~\cite{li2019evaluating,li2018tartan,wang2016interconnect}, and recent RL systems extend this to placing computation---models and parallelism---across heterogeneous GPUs (HetRL~\cite{he2026hetrl}).
RL exacerbates these costs because transient experiences repeatedly traverse the actor\,$\rightarrow$\,Experience Buffer\,$\rightarrow$\,learner path; \sys targets this path with capacity-constrained, bandwidth-aware Experience Buffer placement.

\mypar{Summary}
Across these lines of work, the experience path remains embedded in framework execution or behind passive services; \sys instead exposes it as an explicit, independently optimizable runtime surface.

\section{Conclusion}
\label{sec:conclusion}
We presented \sys, a framework-agnostic runtime that makes the experience path an independently optimizable stage on the actor\,$\rightarrow$\,Experience Buffer\,$\rightarrow$\,learner path.
\sys realizes the \emph{Experience Data Plane} (EDP), exposing \emph{experience ingestion}, \emph{placement}, and \emph{delivery} as explicit runtime control points that run asynchronously while preserving on-policy freshness and off-policy replay semantics.
Guided by measured bandwidths and a compact latency model, it applies capacity-constrained, bandwidth-aware placement and latency-aware scheduling, reducing exposed experience-path latency and improving end-to-end training across distributed RL workloads.

%% file: sections/appendix.tex
\section{Using \sys{} for LLM Post-Training}
\label{app:llm-post-training}

\sys targets traditional distributed RL, but its EDP is defined over the experience path and is independent of framework execution logic. We show that it also generalizes to RL-based LLM post-training by integrating it into \texttt{Verl}~\cite{sheng2025hybridflow}.

\mypar{Ingestion and delivery costs}
In LLM post-training, the work that turns rollouts into training-ready batches is substantial: each sample requires reward (and, for actor-critic methods, value) computation~\cite{sheng2025hybridflow,zhong2025optimizing}, and batches are variable-length sequences carrying auxiliary fields (\eg rewards and masks)~\cite{hu2025openrlhf,ouyang2022training}. \sys abstracts this work as its two experience-path operators, \emph{ingestion} and \emph{delivery}, both of which lie on the critical path.

\mypar{Integration with \texttt{Verl}}
\texttt{Verl}~\cite{sheng2025hybridflow} is a state-of-the-art RLHF framework whose per-iteration dataflow forms an experience path: each iteration generates rollouts and then turns them---through log-prob, reward, and advantage computation---into the training batch consumed by the update. \sys maps this post-generation processing onto \emph{experience ingestion} and the hand-off of the training batch to the learner onto \emph{experience delivery} (\T\ref{tab:verl-map}). By exposing these as independently schedulable operators, the EDP lets \sys overlap ingestion work that \texttt{Verl} otherwise runs serially on the critical path: here, reward depends only on the generated sequences, so \sys runs it concurrently with the remaining post-generation processing rather than afterward, without changing \texttt{Verl}'s reward logic, loss, or compute engines. \looseness=-1

\mypar{Setup and results}
We evaluate this integration by running GRPO~\cite{shao2024deepseekmath} on the \textit{DAPO-Math-17k}~\cite{yu2025dapo} dataset with the \textit{Qwen2.5-Math-7B}~\cite{qwen} model for the task of generating Python code to solve math problems on an 8-GPU NVIDIA A100 node (\S\ref{subsec:exp_setting}).
Integrating \sys reduces exposed experience-path latency from 42\,s to 20\,s per iteration (52\% reduction), which translates into a 4\% end-to-end reduction, as LLM rollout and gradient update dominate iteration time; at industrial post-training scale, even single-digit per-iteration savings accumulate into substantial wall-clock and cost savings.
\looseness=-1
\section{Using \sys{} on \texttt{SRL}}
\label{app:srl}

We further verify this by integrating \sys into \texttt{SRL}~\cite{mei2023srl}.

\mypar{Integration with \texttt{SRL}}
\texttt{SRL}~\cite{mei2023srl} is a state-of-the-art distributed RL framework that splits RL training across actor, policy, and trainer workers operating concurrently via streaming dataflows.
Its experience-path, however, remains embedded: it resides in pinned CPU memory and runs reactively with a fixed one-step overlap (\T\ref{tab:existing_buffers}).
\sys drop-in replaces it: \texttt{SRL}'s workers retain their streaming control flow, while \sys takes over experience placement and the timing and granularity of ingestion and delivery.

\mypar{Setup and results}
We run off-policy DQN on \textit{Qbert}~\cite{mnih2013playing} and on-policy PPO on \textit{Meta-World}~\cite{yu2020meta} on the NVIDIA A100 cluster (\S\ref{subsec:exp_setting}), with eight actors and eight learners; each learner uses \texttt{SRL}'s default policy network.
\T\ref{tab:srl} reports per-iteration latency.
Integrating \sys reduces exposed experience-path latency by 83\% for DQN and 63\% for PPO, translating into 31\% and 18\% lower end-to-end iteration latency, respectively.
We observe a similar off-policy vs.\ on-policy trend as in \texttt{RLlib} (\S\ref{subsec:exp_overall}): replay permits batching and cross-iteration overlap, so DQN sees larger end-to-end gains; for PPO, rollout and update dominate iteration time to begin with, and freshness constraints further limit how much of the remaining experience-path latency can be hidden.

\begin{table}[t!]
	\small
	\centering
	\caption{\sys on \texttt{SRL}: per-iteration end-to-end and exposed experience-path latency (ms).}
	\label{tab:srl}
	\setlength{\tabcolsep}{4.5pt}
	\begin{tabular}{@{}lcccc@{}}
		\toprule
		& \multicolumn{2}{c}{End-to-end} & \multicolumn{2}{c}{Exp-path} \\
		\cmidrule(lr){2-3} \cmidrule(l){4-5}
		Workload & \texttt{SRL} & +\sys & \texttt{SRL} & +\sys \\
		\midrule
		DQN--Qbert & 275.8 & 191.3 ($\downarrow$31\%) & 106.6 & 18.6 ($\downarrow$83\%) \\
		PPO--Meta-World & 235.9 & 193.6 ($\downarrow$18\%) & 63.5 & 23.2 ($\downarrow$63\%) \\
		\bottomrule
	\end{tabular}
\end{table}

\section{Correctness Invariants}
\label{app:correctness}

\sys preserves RL training correctness by maintaining two invariants.

\noindent\textbf{Invariant 1: Training-data integrity.}
\sys preserves the semantic integrity of experience tuples along the actor\,$\rightarrow$\,Experience Buffer\,$\rightarrow$\,learner path.
\textbf{(1)} The Experience Data Plane introduces lightweight boundaries but does not change contents; it only stages data for decoupled execution (\S\ref{sec:decouple}).
\textbf{(2)} Bandwidth-aware placement changes only \emph{where} experiences reside (CPU/GPU, local/remote), not \emph{what} they contain, so tuple semantics and batch composition are preserved (\S\ref{sec:placement}).
\textbf{(3)} Any processing during experience ingestion and experience delivery is algorithm-defined and registered through \sys's interface (\S\ref{sec:overview}); \sys schedules and executes these functions without altering their behavior.

\noindent\textbf{Invariant 2: Algorithm-specific freshness constraints.}
\sys enforces the freshness and replay semantics required by the RL algorithm.
For \emph{on-policy} training, updates must consume samples from the current policy iteration, so \sys allows only within-iteration mini-batch pipelining and excludes cross-iteration reuse by construction (\S\ref{sec:decouple}).
For \emph{off-policy} training, replay across iterations is permitted, so \sys may batch and prefetch samples while bounding staleness through the EDP configurations (\S\ref{sec:decouple}). Replay state such as priorities~\cite{schaul2015prioritized} is updated by algorithm-defined callbacks (Invariant~1(3)), whose buffer dependencies \sys leaves intact.
In all cases, the scheduler chooses overlap and granularity only from these feasible spaces, ensuring optimization never violates correctness (\S\ref{sec:scheduling}).

\begin{table}[t!]
	\small
	\centering
	\caption{\sys mapped onto \texttt{Verl}'s per-iteration experience path (GRPO): post-generation processing (log-prob, reward, advantage) is \emph{ingestion}; the hand-off of the training batch to the update workers is \emph{delivery}.}
	\label{tab:verl-map}
	\setlength{\tabcolsep}{4pt}
	\begin{tabular}{@{}clcl@{}}
		\toprule
		\# & \texttt{Verl} stage & \multicolumn{2}{c}{Role} \\
		\midrule
		1 & \texttt{generate\_sequences} (rollout) & \multicolumn{2}{l}{actor (upstream)} \\
		\midrule
		2 & old\,/\,ref log-prob & \multirow{4}{*}{\textbf{EDP}} & \multirow{3}{*}{ingestion} \\
		3 & reward, KL penalty & & \\
		4 & \texttt{compute\_advantage} & & \\
		\cmidrule(l){4-4}
		5 & dispatch batch to update workers & & delivery \\
		\midrule
		6 & \texttt{update\_actor} & \multicolumn{2}{l}{learner (downstream)} \\
		\bottomrule
	\end{tabular}
\end{table}

\section{Placement-Migration Details}
\label{app:migration}

We detail the online placement migration introduced in \S\ref{subsec:online_migration}: the runtime procedure (\A\ref{alg:migration}) and the transfer-plan formulation that minimizes migration makespan.

During the offline profiling phase, \sys sweeps a discretized range of per-sample sizes and solves the placement scan once per point, producing a compact placement map $\mathcal{M}$.
At runtime, migration resolves $\mathcal{M}(x)$, briefly pauses ingestion and delivery, transfers data to the new placement, and resumes under the updated placement.
The EDP boundaries keep actors and learners decoupled while the buffer reallocates.

A placement specifies a device set and the data volume held on each device.
Migrating from placement $p$ (device set $D$) to $p'$ (device set $D'$) requires deciding how much data each source device sends to each target device.
Let $f_{ij}$ denote the volume sent from $d_i \!\in\! D$ to $d'_j \!\in\! D'$; since transfers execute in parallel on dedicated streams, the migration makespan is the slowest individual transfer.
\sys minimizes this makespan:
\begin{equation}
	\label{equ:migration}
	\min_{f_{ij} \ge 0} \max_{i,j} \frac{f_{ij}}{BW(d_i \!\to\! d'_j)}
	~~~~\text{s.t.}~~
	\textstyle\sum_j f_{ij} = V_i,~~
	\textstyle\sum_i f_{ij} = V'_j,
\end{equation}
where $V_i$ is the data volume on $d_i$ and $V'_j$ is the required volume on $d'_j$.
Since the placement map has only a small number of regions, \sys pre-computes transfer plans for the expected transitions and reuses them online.

\section{Iteration-Latency Model Details}
\label{app:latency-model}

We expand the iteration-latency model introduced in \S\ref{subsec:latency}, making explicit the hidden-latency terms behind its compact exposed-latency form.
For ingestion and delivery, overlap can hide at most the shorter of the neighboring compute stage and the per-unit operator cost:
\begin{align}
hide_{\text{ing}} &=
\gamma(g_{\text{ing}})\cdot
\min\!\bigl(T_{\text{rollout}}, \bar{T}_{\text{ing}}(g_{\text{ing}})\bigr), \\
hide_{\text{del}} &=
\gamma(g_{\text{del}})\cdot
\min\!\bigl(T_{\text{update}}, \bar{T}_{\text{del}}(g_{\text{del}})\bigr).
\end{align}
The factor $\gamma(g)$ captures how much of that bound is practically overlapable:
\begin{equation}
\gamma(g)=
\begin{cases}
1, & g \ge 1 \\
1-g, & 0<g<1 .
\end{cases}
\end{equation}
Thus, in off-policy batching ($g\ge 1$), overlap can hide the full per-unit operator cost, while in on-policy fractional pipelining ($0<g<1$), only the unexposed fraction scales with $(1-g)$.

\begin{algorithm}[t]
	\begin{footnotesize}
		\caption{\small Online placement migration.}
		\label{alg:migration}
		\KwIn{Placement map $\mathcal{M}$, pre-solved transfer plans $\mathcal{F}$, active placement $p^*$, per-sample size $x$.}
		\KwOut{Updated active placement $p^*$}
		$p^*_{\text{new}} \gets \mathcal{M}(x)$ \tcp*{$\mathcal{O}(\log R)$ lookup}
		\textbf{If} $p^*_{\text{new}} = p^*$ \textbf{then} \Return $p^*$ \tcp*{no migration needed}
		$\{f_{ij}\} \gets \mathcal{F}(p^*,\, p^*_{\text{new}})$ \tcp*{pre-computed plan}
		\textit{Pause} ingestion and delivery operators\;
		\ForEach{$(d_i, d'_j)$ with $f_{ij} > 0$ \textbf{in parallel}}{
			Transfer $f_{ij}$ data: $d_i \!\to\! d'_j$ on dedicated stream\;
		}
		$p^* \gets p^*_{\text{new}}$\;
		\textit{Resume} ingestion and delivery under $p^*$\;
		\Return $p^*$\;
	\end{footnotesize}
\end{algorithm}

\section{Cost-Model Accuracy}
\label{app:costmodel}

\sys's optimizer selects configurations by argmin over cost-model predictions, so rank fidelity matters more than pointwise accuracy.
We evaluate both on the $224$ manual configurations of \F\ref{fig:auto_vs_manual} (seven workload dimensions $\times$ $32$ configurations each), predicting each configuration's $T_{\text{iter}}$ with the model of \S\ref{subsec:latency} and comparing against measurement (\T\ref{tab:costmodel}).
The comparison here uses $T_{\text{iter}}$, the optimizer's objective in Eq.~\eqref{eq:sched_opt}, rather than the exposed experience-path latency plotted in \F\ref{fig:auto_vs_manual}.
Pointwise, predictions follow measurements with near-perfect linear correlation (Pearson $r{=}0.990$) and low absolute error (MAPE $20.8\%$).
Rank-wise, the model is stronger still: for every dimension, the predicted ordering of the eight lowest-latency configurations exactly matches the measured ordering, so the optimizer always picks the truly best configuration.

\begin{table}[t]
    \small
    \centering
    \caption{Cost-model accuracy per workload dimension $d$ of \F\ref{fig:auto_vs_manual} ($32$ configurations each). For every dimension, the predicted ordering of the top-$8$ configurations exactly matches the measured ordering.}
    \label{tab:costmodel}
    \setlength{\tabcolsep}{4.5pt}
    \begin{tabular}{@{}rrrc@{}}
        \toprule
        & \multicolumn{2}{c}{Best $T_{\text{iter}}$ (ms)} & \\
        \cmidrule(lr){2-3}
        $d$ & Measured & Predicted & Top-$8$ ranking match \\
        \midrule
        $128$       & $17.1$ & $20.1$ & \checkmark \\
        $256$       & $18.3$ & $20.8$ & \checkmark \\
        $512$       & $25.4$ & $29.9$ & \checkmark \\
        $1{,}000$   & $26.4$ & $33.9$ & \checkmark \\
        $2{,}000$   & $30.0$ & $35.8$ & \checkmark \\
        $5{,}000$   & $42.6$ & $43.2$ & \checkmark \\
        $10{,}000$  & $48.6$ & $58.9$ & \checkmark \\
        \bottomrule
    \end{tabular}
\end{table}